\documentclass[twocolumn,prl]{revtex4-1}
\usepackage{amsthm,setspace,afterpage}
\usepackage[usenames,dvipsnames]{xcolor}
\usepackage{amsmath,amssymb,graphicx,bm,subfigure}
\usepackage{hyperref}

\graphicspath{ {./figures/} }

\begin{document}
\title{Landau Theory and the Emergence of Chirality in Viral Capsids}
\author{Sanjay Dharmavaram$^{1}$, Fangming Xie$^{4}$, William Klug$^{1}$, Joseph Rudnick$^{2}$, Robijn Bruinsma$^{2,3}$}
\email{sanjaydm@ucla.edu}
\affiliation{$^{2}$Department of Physics and Astronomy,}
\affiliation{$^{1}$Department of Mechanical and Aerospace Engineering,}
\affiliation{$^{3}$Department of Chemistry and Biochemistry, 
  University of California, Los Angeles,  CA 90095, USA}
\affiliation{$^4$Department of Physics, University of Science and Technology of China, Hefei, Anhui, China}
\begin{abstract}
We present a generalized Landau-Brazovskii theory for the solidification of chiral molecules on a spherical surface. With increasing sphere radius one encounters first intervals where robust achiral density modulations appear with icosahedral symmetry via first-order transitions. Next, one encounters intervals where fragile but stable icosahedral structures still can be constructed but \textit{only} by superposition of multiple irreducible representations. Chiral icoshedral structures appear via continuous or very weakly first-order transitions. Outside these parameter intervals, icosahedral symmetry is broken along a three-fold axis or a five-fold axis. The predictions of the theory are compared with recent numerical simulations.
\end{abstract}
\maketitle

Landau theory is a central method for the analysis of phase transitions and spontaneous symmetry breaking. When complex fluids solidify, the emergent density modulations often adopt a variety of contending symmetries. Landau theory has been a powerful tool for the analysis of the phase diagrams \cite{Kats}. In this letter, the Landau theory of the solidification of chiral molecules confined to a spherical surface is applied to the solidification of viral capsids \cite{Garmann, Bruinsma}. Capsids are spherical protein shells that surround viral genome molecules. Spherical viral capsids usually -- though not always -- have icosahedral symmetry \cite{caspar}. On the other hand, numerical studies of systems of interacting particles confined to spherical surfaces find that icosahedral symmetry must contend with a variety of other symmetries \footnote{J. M. Voogd, PhD thesis, Universiteit van Amsterdam, 1994},\cite{Fejer, Zandi, Wales, Paquay}. 

According to Landau theory, the density modulation of the ordered phase on a spherical surface should transform under the symmetry group $G_0$ of the isotropic fluid phase according to a single irreducible representation (``irrep") of $G_0$. The symmetry group $G$ of the ordered phase must be an isotropy subgroup of $G_0$. Because the capsid proteins, or ``subunits", are chiral molecules (formed from the L-isomers of amino-acids) $G_0$ should be the chiral group SO(3) of rotations without inversion. The density modulation of the ordered phase must then be expandable in the basis set of the $2l+1$ spherical harmonics ${Y}_l^m$ for a particular $l$ value, itself determined by minimization of the Landau free energy. As shown in refs.~\cite{lorman2008landau, lorman2007density}, if one \textit{demands} that the ordered phase has chiral icosahedral symmetry, so with $G$ the chiral icosahedral group $I$, then the expansion coefficients can be obtained from group theory. This leads to the icosahedral spherical harmonics $\mathcal{Y}_h(l)$\cite{Klein}. Irreps of SO(3) with icosahedral symmetry can be constructed from even spherical harmonics for $l=6j+10k$ with $j,k\in\{0,1,2,\cdots\}$ and from odd spherical harmonics for $l=15+6j+10k$ \cite{golubitsky2012singularities}. If $I$ is the isotropy subgroup then only odd values of $l$ are allowed. This means that the smallest icosahedral capsids \cite{caspar} should have surface density modulations proportional to $\mathcal{Y}_h(l=15)$ \cite{lorman2008landau, lorman2007density} (maxima and minima of the density modulation correspond here to the locations of subunits, whose density is odd under inversion). 

The cubic non-linear term in the free energy density integrates to zero for odd $l$, which means that the solidification transition should be a \textit{continuous transition}. This is surprising since solidification normally is a first-order transition \cite{Kats}. This suppression of the first-order activation barrier by chirality would have important consequences for viral assembly. Models for protein-by-protein capsid assembly under equilibrium conditions encounter activation energy barriers that can be very large compared to the thermal energy \cite{Zandi2}. If the solidification transition for chiral molecules on a spherical surface indeed has no activation barrier, then this could be an important argument in favor of collective descriptions of capsid assembly \cite{Garmann, Bruinsma, Hagan} as opposed to the protein-by-protein assembly models.

In this letter we re-examine the Landau theory for solidification on a spherical surface of molecules with chiral interactions \cite{lorman2008landau, lorman2007density} now without requiring the density modulation to have icosahedral symmetry and without requiring it to be odd under inversion. Following other Landau theories of chiral complex liquids \cite{deGennes, Helfrich}, chirality is introduced here by including the lowest-order non-zero pseudo-scalar invariant of the order parameter in the free energy. The Landau-Brazovskii (LB) free energy density \cite{brazovskii1975phase, Kats} for solidification generalized to chiral systems is then
\begin{equation}
\begin{split}
  F=&\int_{S^2}\bigg[\left(\frac{1}{2}\Big( (\Delta+k_o^2)\rho\Big)^2+\frac{r}{2}\thinspace\rho^2+\frac{u}{3}\thinspace\rho^3+\frac{v}{4}\thinspace\rho^4+....\right)\\&
+\chi \bigg(\nabla \rho\cdot(\nabla\nabla\rho)^2\cdot(\mathbf{n}\times \nabla\rho)\bigg)\bigg]\;dS.
  \label{eq:LBEnergy}
\end{split}
\end{equation}
The integral is over a spherical surface $S^2$. The first term is the standard LB free energy density.  $\Delta$ is the Laplace-Beltrami operator on a spherical surface and $k_0$ the ratio of the radius of the sphere and the characteristic length scale of the density modulation. Lengths are measured in units of the sphere radius. The second part of the first term is an expansion of the free energy density in powers of the scalar invariants of the density modulation $\rho$ with $r$, $u$, and $v$ the lowest-order expansion coefficients ($u$ is negative and $v$ is positive).  The second term of Eq.\ref{eq:LBEnergy} is the lowest-order pseudo-scalar that can be constructed from $\rho$ whose integral over a closed surface is non-zero \footnote{This term, which is similar to the chiral term of the Helfrich-Prost theory for curved chiral surfaces \cite{Helfrich}, is further discussed in section S1 of supplementary material.}. Here, $\mathbf{n}$ is a unit vector normal to the sphere surface and $\chi$ is a parameter that measures the strength of chiral symmetry breaking \footnote{To show that the uniform phase of this free energy has SO(3) symmetry  rather than O(3) it is necessary to compute higher order density correlation function in the presence of thermal fluctuations}. 

Since the spherical harmonics $Y_l^m$ are eigenfunctions of the $\Delta$ operator with eigenvalue $-l(l+1)$, one can divide up the $k_0$ axis into segments $l^2<{k_0}^2<(l+1)^2$ so that in each segment the quadratic term in the free energy for that particular value of $l$ first changes sign as one decreases the control parameter $r$ \footnote{The stability threshold is $r_l(k_0)=-\left(k_0^2-l(l+1)\right)^2$}. The expansion coefficients of the density in the basis of $Y_l^m$ are obtained by minimizing the full free energy, which results in $2l+1$ coupled non-linear equations $\langle Y_l^m|\frac{\partial F(\rho)}{\partial\rho}\rangle=0$ \footnote{The explicit form of the equations is provided in section S2 of supplementary material.}. For a particular solution of these equations to be stable, the eigenvalues of the Hessian matrix of second derivatives for that solution must be positive, apart from three eigenvalues that are zero associated with global rotation \footnote{Hessian matrices were diagonalized numerically using Mathematica}. The first segment along the $k_0$ axis that supports a non-trivial icosahedral subgroup is the one for $l=6$. The 13 coupled equations $\langle Y_6^m|\frac{\partial F(\rho)}{\partial\rho}\rangle=0$ are solved by $\rho_6=\xi  \mathcal{Y}_h(6)$. Here
\begin{equation}
  \mathcal{Y}_h(6) = Y_6^0+\sqrt{\frac{7}{11}}(Y_6^5-Y_6^{-5}).
\label{ish}
\end{equation}
is the first non-trivial icosahedral spherical harmonic. The order parameter amplitude $\xi$ obeys the cubic equation
\begin{equation}
  r\xi+\tilde{u}\xi^2+\tilde{w}\xi^3=0,
  \label{FO}
\end{equation}
with $\tilde{u}=u(50\sqrt{13}/323\sqrt{\pi})$ and $\tilde{w}=2145w/(1564\pi)$. Equation \ref{FO} has the standard form of a first-order phase transition. The relevant eigenvalues of the $13\times13$ Hessian matrix of the $\rho_6$ solution branch are positive when the uniform state is unstable \footnote{see last paragraph of section S2 of supplementary material}. The coefficient $\chi$ does not enter in Eq. \ref{FO} because the pseudoscalar invariant is zero in \textit{any} of the $2l+1$ dimensional subspaces \footnote{see section S3 of supplementary material}. Similar first-order solidification transitions to a stable icosahedral state take place in the subsequent $l=10$ and $l=12$ segments. If one identifies maxima of the density modulation with ``capsomers" (groups of 5 or 6 subunits) rather then subunits themselves \cite{Zandi} then $l=6$ corresponds to a $T=1$ capsid and $l=10$ and $l=12$ to $T=3$, respectively, $T=4$ capsids \footnote{The corresponding density modulations are shown in section S4 of the supplementary material}. 

The next icosahedral intervals are for $l$=15, 16, 20 and 22. The icosahedral harmonics remain solutions of the projection equations $\langle Y_l^m|\frac{\partial F(\rho)}{\partial\rho}\rangle=0$ but their Hessian now has negative eigenvalues \footnote{It is shown analytically in section S5 of supplementary material that $\mathcal{Y}_h(15)$ is unstable for any value of $u$ and $v$}. Stable icosahedral states that involve these icosahedral harmonics still can be constructed but only by going beyond conventional Landau theory and include multiple irreps of SO(3). As an example, in the enlarged product space of the $l=15$ and $ l=16$ subspaces (denoted by $``15\times16"$) icosahedral density modulations would have the form
\begin{equation}
  \rho(\zeta,\xi) = \zeta \mathcal{Y}_h(15) + \xi \mathcal{Y}_h (16),
  \label{eq:coupled_ansatz}
\end{equation}
The free energy has an extremum when the two-component order-parameter $(\zeta,\xi)$ obeys a pair of coupled cubic equations \footnote{ $c_{16}=(k_0^2-16\times17)^2$, $c_{15}=(k_0^2-15\times16)^2$, $a_1/16=-80.4$, $a_2/16=-3084.1$, $u_1=0.13494 u$, $u_2=0.234946 u$, $v_1=1.04204 w$, $v2=0.681228 w$, $u_3=0.623577 u$, $v_3=0.789107 w$, and $v_4=0.904033 w$}:
\begin{subequations}
  \begin{equation}
    (c_{16}+r)\xi + u_1 \xi^2 + u_2\zeta^2 + v_1 \xi^3+ v_2 \xi\zeta^2 - \chi(3a_1 \xi^2 \zeta+a_2\zeta^3)= 0,
    \label{xi}
  \end{equation}
  \begin{equation}
    (c_{15}+r)\zeta + u_3\xi\zeta + v_3 \zeta^3 + v_4\zeta \xi^2 -\chi(a_1 \xi^3 +3a_2\zeta^2\xi)= 0,
    \label{zeta}
  \end{equation}
  \label{zeta}
\end{subequations}
Numerical solution at the midpoint $k_0=16$ between the $l=15$ and $l=16$ stability intervals shows that in that case the two components $(\zeta,\xi)$ of the resulting order parameter are nearly proportional to each other and comparable in magnitude \footnote{The resulting density modulations are shown in section S6 of the supplementary material and compared with the result of a two-state MC simulation of T=7 CK capsids composed of 72 capsomers \cite{Zandi}}. The chiral term in the free energy does not vanish in the enlarged $15\times16$ subspace and now enters in these equations. For $\chi=0$, there are two degenerate solutions, which we denote by D and L, that transform into each other under inversion (with $\zeta\rightarrow-\zeta$). 
For $\chi\neq0$ but small, the free energy of the D and L branches shifts by an amount $\Delta F_c =\chi\left(a_1\xi_0^3\zeta_0+a_2\xi_0\zeta_0^3\right)$ where ($\zeta_0,\xi_0$) is the $\chi=0$ solution \footnote{There is also a shift of the two branches by a term that is even in $\zeta$}. Since $\Delta F_c$ is odd in $\zeta_0$ the degeneracy between the D and L branches under inversion is lifted by an amount proportional to $\chi$. In the capsomer interpretation, the $15\times16$ superposition state corresponds to the -- intrinsically chiral -- T=7 shell while it correspond in the subunit interpretation to the T=1 shell. In the latter case, the individual subunits are now neither odd nor even under inversion as is in fact the case for actual chiral molecules \footnote{This is illustrated in Fig. S4 of the supplementary material section S6. Similar results hold for larger values of $l$. For example, the $21\times22$ branch produces a chiral T=3 shell in the subunt picture and a T=13 shell in the capsomer picture. The $20\times21$ branch produces a chiral T=2 non-CK icosahedral shell.}. 

\begin{figure}
 \includegraphics[width=3.5in]{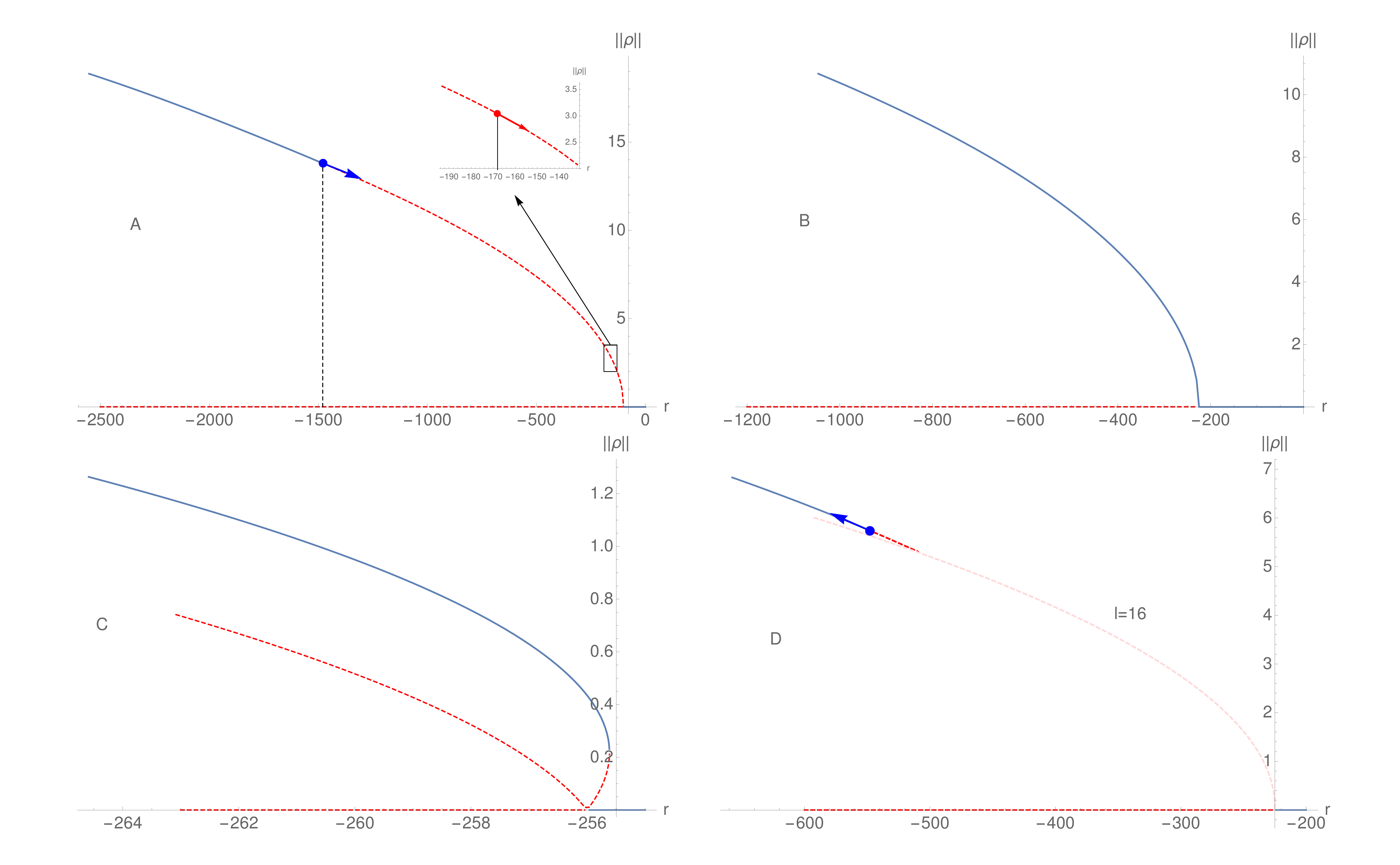}
 \caption{Stability diagrams of the 15$\times$16 chiral icosahedral states for different values of the dimensionless radius $k_0$. Horizontal axis: control parameter $r$. Vertical axis: order parameter amplitude $||\rho||=(\xi^2+\zeta^2)^{1/2}$. Solid blue: stable section. Dashed red: unstable section.  (A): $k_0\simeq15.8$ Blue dot: instability point of the icosahedral state against irrep $3$. Red dot:  instability points against irrep $4$. (B): $k_0\simeq15.9$, (C): $k_0=16$, (D): $k_0\simeq16.5$. The unstable section is now mostly a pure $l=16$ state.}
\label{BD}
\end{figure}

\begin{figure}
\includegraphics[width=3in]{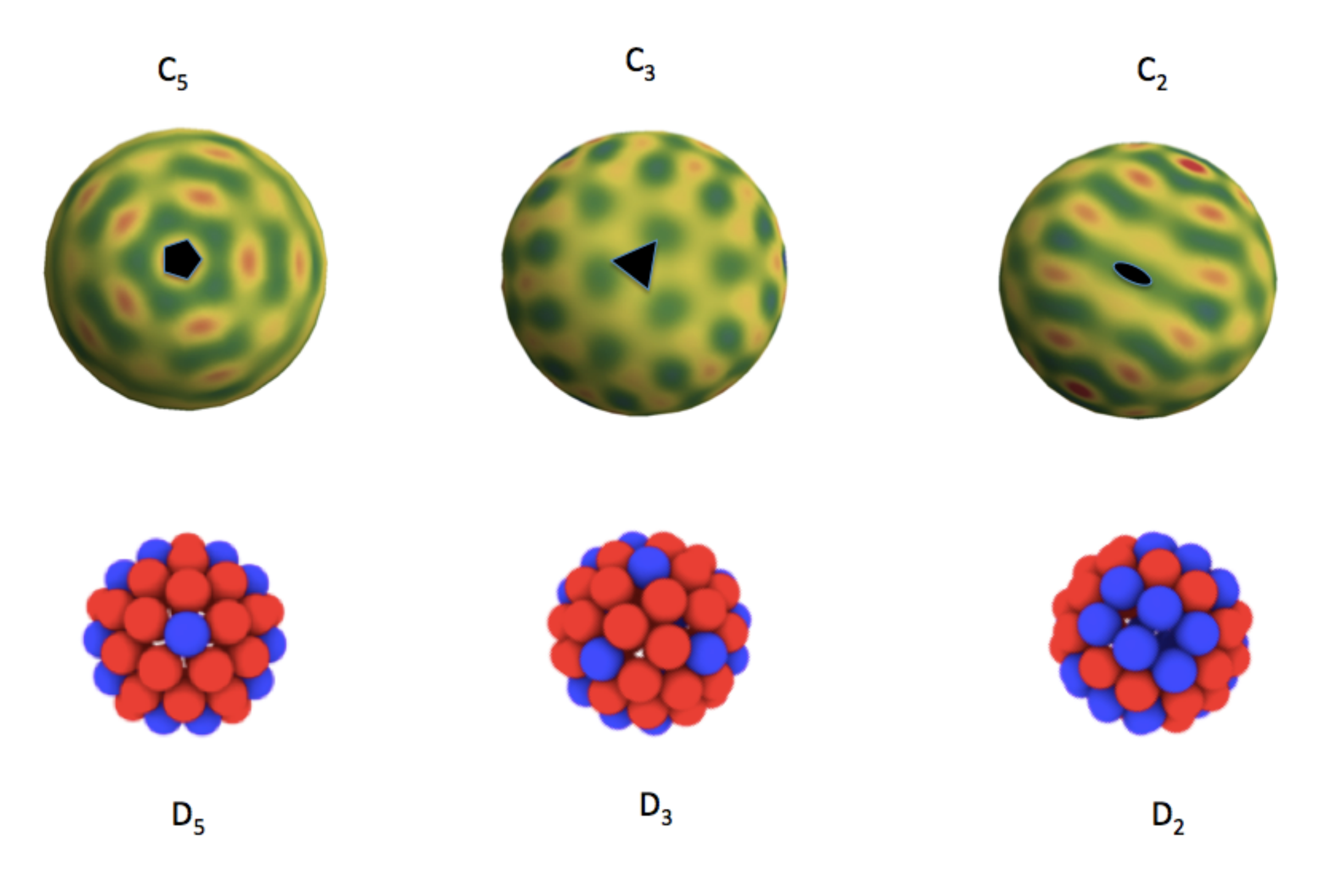}
 \caption{(top) Three unstable eigenvectors of the Hessian of the 15$\times$16 chiral icosahedral states with, respectively, $C_5$, $C_3$, and $C_2$ rotational symmetry (left to right) . (bottom) Competing groundstates of 72 one-state Lennard-Jones particles on a sphere with (left to right) $D_5$, $D_3$ and $D_2$ symmetry. Blue spheres: 5-fold coordinated  particles. Red spheres: 6-fold coordinated particles. From ref.\cite{Paquay} with permission.}
\label{BD2}
\end{figure}

\begin{figure}
\includegraphics[width=2in]{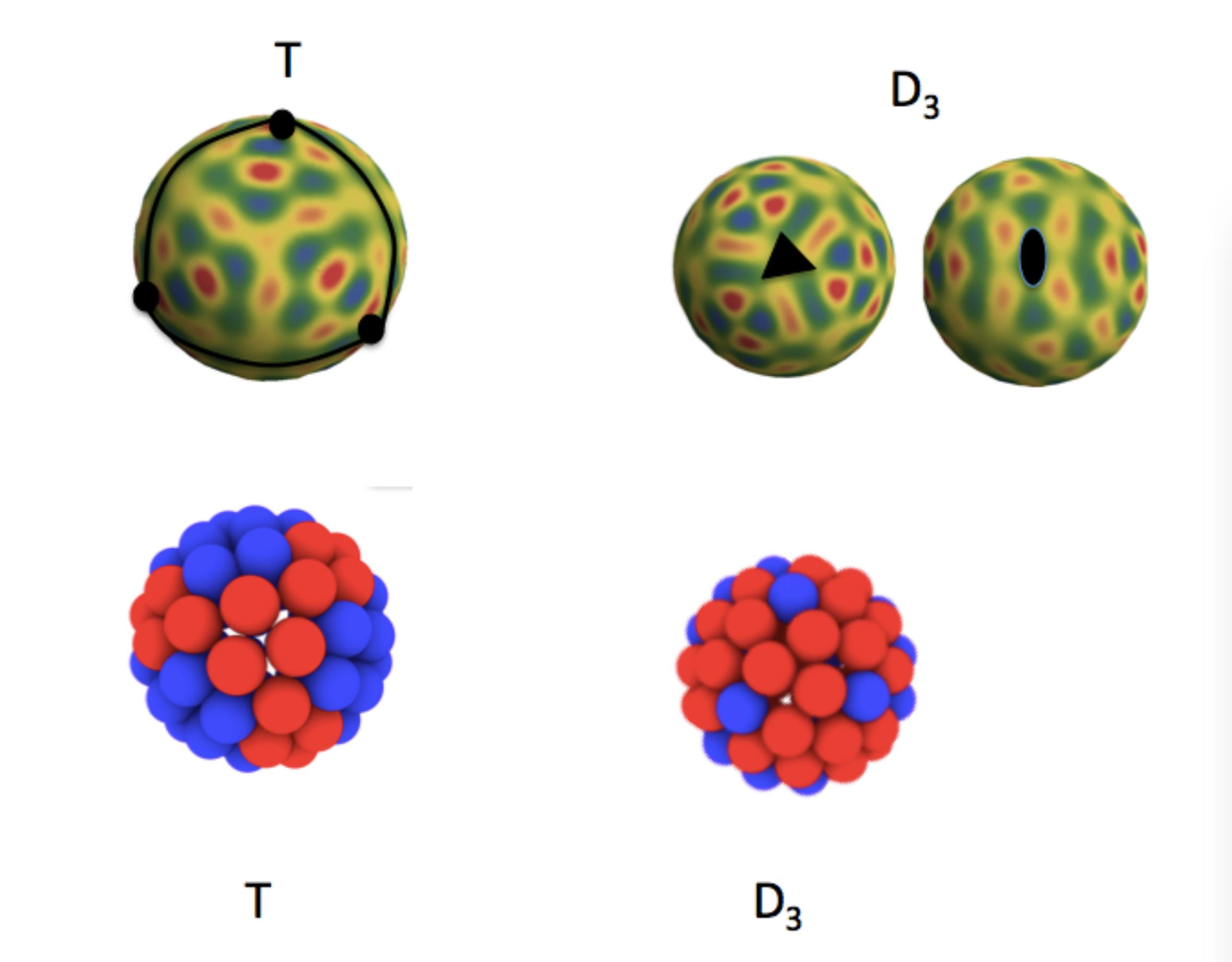}
 \caption{(top) Unstable eigenvectors with tetrahedral and $D_3$ symmetry. (bottom) Tetrahedral and $D_3$ groundstates of 72 Lennard-Jones particles on a sphere. The two and three fold axes are perpendicular for $D_3$. From ref.\cite{Paquay} with permission.}
\label{tetra}
\end{figure}

The stability of the superposition state is a sensitive function of the dimensionless radius $k_0$. Increasing the $k_0$ parameter from the stability center of the $l=15$ segment at $k_0\simeq15.4$ towards the stability center of the $l=16$ segment at $k_0\simeq16.5$, one finds that for decreasing $r$ the uniform state initially does not transform directly into an icosahedral state. Instead, a stable superposition state appear only for values of $r$ that are well below the stability limit of the uniform phase, as shown in Fig.\ref{BD} A.
 Starting in this stable icosahedral state, and now increasing $r$, groups of three-fold degenerate eigenvalues $\lambda_3\propto (r-r^*(k_0))$ change sign at a first threshold $r=r^*(k_0)$ (blue dot). Three corresponding eigenvectors of the Hessian with 2-fold, 3-fold, and 5-fold axial symmetry are shown in Fig.\ref{BD2}.
For $r$ close to the stability limit of the uniform phase, one encounters a second instability point along the now unstable $15\times16$ branch (red dot), where groups of four degenerate eigenvalues change sign. The corresponding eigenvectors have tetrahedral or $D_3$ symmetry (see Fig.\ref{tetra}).
As $k_0$ increases towards the boundary point $k_0=16$, both instability points slide down the stability curve and disappear around $k_0=15.9$. Over a short interval of $k_0$ values, between 15.90 to 15.96 a stable $15\times16$ branch appears directly from the uniform phase. Remarkably, within our numerical precision the transition indeed is continuous as shown in Fig.\ref{BD} (B). For $k_0$ slightly larger than 15.96. There still is a direct transition into the $15\times16$ branch but the transition is now weakly first order (see Fig.\ref{BD} C). As $k_0$ increases towards the center of the $l=16$ stability segment at $k_0\simeq 16.5$, the icosahedral state again becomes progressively less stable as shown in Fig.\ref{BD} D \footnote{The instability sequence is not symmetric around $k_0=16$: the unstable icosahdral branch now mostly is a pure $l=16$ icosahedral harmonic with the $15\times16$ state only appearing close to the stability limit of the icosahedral state}.

The transition from the icosahedral state to one with lower symmetry can be analyzed by classical Landau theory. The symmetry group $G_0$ is now the chiral icosahedral group $I$, which has one one-dimensional (``1D") irrep, two 3D irreps, a 4D and a 5D irrep \cite{Hoyle}. The first instability in Fig.\ref{BD} (A) is associated with one of the two 3D irreps, denoted by 3, while the second is associated with the 4D irrep. The isotropy subgroups of $I$ associated with the 3D irrep 3 are $C_5$, $C_3$, and $C_2$. A natural orthogonal coordinate system in the 3D space of irrep 3 is one with the $(1,1,1)$ direction along a three-fold symmetry direction, the $(0,0,1)$ direction along a two-fold direction \footnote{The direction $\eta_1=0$ and $\eta_2^2 = 1/2(3-5^{1/2})\eta_3^2$ is then along a 5-fold axis}. The order parameter is then a 3D vector $\vec{\eta}=(\eta_1,\eta_2,\eta_3)$ in this space. Using symmetry arguments \footnote{See section S7 of Supplementary Material supplementary material.} the Landau energy $\delta F(\vec{\eta})$ can be shown to have the form:
\begin{equation}
\begin{split}
&\delta F(\vec{\eta}) \propto \frac{\lambda_3}{2}|\eta|^2+\frac{V}{4}|\eta|^4+\frac{W}{6}|\eta|^6+\Delta\frac{\sqrt{5}}{2}\eta_1^2\eta_2^2\eta_3^2\\&+\Delta\frac{\sqrt{5}}{60}(\eta_1^6+\eta_2^6+\eta_3^6)+\frac{\Delta}{4}\left(\eta_1^4(\eta_3^2-\eta_2^2)+cyclic\medspace perm.\right)
\label{ISM}
\end{split}
\end{equation}
to sixth order in $\eta$. Here, $\lambda_3$ is the eigenvalue that changes sign at the transition while $V$ and $W$ are positive constants \footnote{The terms proportional to $W$ and $\Delta$ in Eq.\ref{ISM} are generated by sixth and higher order terms in $\rho$ in the LB free energy density Eq. \ref{eq:LBEnergy}.}. A continuous transition takes place at $\lambda_3=0$. For $\Delta=0$, the Landau energy is isotropic with arbitrary rotations in $\vec{\eta}$ space connecting degenerate states. For $\Delta$ negative, the minimum of $\delta F(\vec{\eta})$ lies along the $C_3$ direction and for $\Delta$ positive along the $C_5$ direction. For negative $\lambda_3$, the system thus has either a three-fold or a five-fold axis. The second instability in Fig.\ref{BD} (top left) is associated with the 4D irrep, which has $T$ and $D_3$ as its isotropy subgroups. The corresponding eigenvectors are shown in Fig.\ref{tetra}. The $\delta F(\eta)$ for the 4D irrep contains cubic invariants in $\eta$. This means that icosahedral symmetry breaking in this case is discontinuous. Finally, an instability driven by the 5D irrep would have $D_2$, $D_3$ and $D_5$ as its isotropy subgroups.

Recently, numerical simulations of the groundstate of 72 ``capsomer particles" on a spherical surface \cite{Zandi, Paquay} --  corresponding to a T=7 shell in the capsomer interpreation -- that interact via Lennard-Jones (LJ) interaction. Depending on parameters, they found an icosahedral groundstate as well as competing groundstates with $D_2$, $D_3$, $D_5$, and $T$ symmetry (see Figs.\ref{BD2} and \ref{tetra}). They did not encounter however the $C_3$ and $C_5$ states. This suggests that for a LJ particle system, icosahedral symmetry is broken by the 4D and 5D irreps rather than the 3D irrep of LB theory. 

What are the implications of these results for viral assembly? In the capsomer picture, a $T=1$ capsid appears via a robust first-order solidification transition and is quite stable. The capsid is chiral only in so far as capsomer particles are chiral. In the subunit picture, the $T=1$ capsid  appears as a mixed $15\times16$ state that is much less stable. It requires careful fine-tuning of the dimensionless radius $k_0$. The ordering transition of the subunit fluid is continuous or very weakly first-order. Chirality is now directly expressed in the subunit density modulation, which is neither even nor odd under inversion. There is apparently no unique Landau description and experiment may have to determine which one is  appropriate. The T=7 bacteriophage virus HK97 \cite{conway} may be an ideal ``testbed" to test the theory. In the presence of high salt, Prohead 1 capsids of HK97 dissociate into interconverting pentamers and hexamers that are quite stable \cite{Xie} so the capsomer picture is appropriate. The kinetics of Prohead 1 reassembly indicates that it is a highly cooperative process with some phase coexistence between capsids and capsomers \cite{Xie}, possibly consistent with the weakly first-order transition picture of Fig.\ref{BD} B. Inducing point mutations on the genes for the capsid proteins could destabilize the fragile icosahedral state and reveal the instabilities of Fig.\ref{BD2}. Next, the theory predicts that intrinsically chiral icosahedral capsids, such as T=7 and T=13, should be very prone to undergo a transition into a lower symmetry state with $C_3$ or $C_5$ symmetry. The capsid should have an \textit{oval deformation} for broken icosahedral symmetry because the corresponding eigenvectors are nearly odd under inversion. This has not yet been reported for HK 97 but capsids of the HIV virus do have an oval shape.

Finally, we note that the necessity of using of multiple irreps becomes less puzzling by noting that in the limit that the sphere radius goes to infinity, the density modulation should reduce to the superposition of three periodic density waves whose wavevectors make an angle of 120 degrees with respect to each other (a hexagonal crystal). The expansion of a plane wave into spherical harmonics requires an infinite series of $l$ terms so an \textit{infinite series of irreps} of SO(3) would be required to describe solidification in the limit of large sphere radii. 

\section{Acknowledgments}
We would like to thank Vladimir Lorman and Alexander Grosberg for helpful discussions, the NSF for support under DMR Grant No. 1006128 and the Aspen Center for Physics for hosting a workshop on the physics of viral assembly.

%

\end{document}


\section{S1: Constructing the Chiral Term}
We construct the chiral contribution to the free energy (the last term
in equation (1)) by following the Landau prescription of expanding
the free energy density as polynomial in $\rho$ and its derivatives,
and demanding the free energy to be invariant under the symmetry of
the high-temperature phase -- proper rotations, in our case.

We begin by first pointing out that instead of working in spherical
coordinates, we treat the problem on a flat plane where the order
parameter $\rho$ is a function of the cartesian coordinates
$(x,y)$. The proper rotations in this case are elements of $SO(2)$
(with the axis of rotation chosen to be normal to the plane). The form
for the free energy that we obtain can then be generalized to the
surface of a sphere by replacing derivatives of $\rho$ by their
appropriate covariant forms.

Consider the following general form for the free energy:
\begin{equation}
  F=\int f(\rho,\nabla\rho,\nabla\nabla\rho)\;dx\;dy,
\label{eq:sup_free energy}
\end{equation}
written in terms of the energy density $f$.

Under orthogonal (proper) rotations, $\mathbf{R}_\theta\in SO(2)$, where $\mathbf{R}_\theta$ represents the $2\times 2$ rotation matrix
$\left(
\begin{array}{cc}
  \cos\theta & -\sin\theta \\
  \sin\theta & \cos\theta
\end{array}
\right),$
if
\begin{equation}
\left(
\begin{array}{c}
  x \\
  y
\end{array}
\right)\mapsto
\left(
\begin{array}{c}
  \tilde{x} \\
  \tilde{y}
\end{array}
\right):=
\mathbf{R}_\theta^T\left(
\begin{array}{c}
  x \\
  y
\end{array}
\right),
\end{equation}
for which $\rho$ transforms according to the prescription,
$$\rho(x,y)\mapsto \rho(\tilde{x},\tilde{y}),$$ it is straightforward
to verify that the derivatives of $\rho$ must transform as:
\begin{equation}
\left(
\begin{array}{c}
  \rho_x \\
  \rho_y
\end{array}
\right)\mapsto \mathbf{R}_\theta\left(
\begin{array}{c}
  \rho_{\tilde x} \\
  \rho_{\tilde y}
\end{array}
\right),
\end{equation}
\begin{equation}
\left(
\begin{array}{cc}
  \rho_{xx} & \rho_{xy} \\
  \rho_{xy} & \rho_{yy}
\end{array}
\right)
\mapsto \mathbf{R}_\theta\left(
\begin{array}{cc}
  \rho_{\tilde{x}\tilde{x}} & \rho_{\tilde{x}\tilde{y}} \\
  \rho_{\tilde{x}\tilde{y}} & \rho_{\tilde{y}\tilde{y}}
\end{array}
\right)\mathbf{R}_\theta^T.
\end{equation}
The free energy (\ref{eq:sup_free energy}) is invariant under $SO(2)$, provided,
\begin{equation}
\int f(\rho,\nabla\rho,\nabla\nabla\rho)\;dxdy = \int
f(\rho,\mathbf{R}_\theta\nabla\rho,\mathbf{R}_\theta\nabla\nabla\rho\mathbf{R}_\theta^T)\;d\tilde{x}d\tilde{y}.
\label{eq:sup_inv_energy}
\end{equation}
Since the second integral (through its dependence on
$\rho$) in the previous equation is a function of
$(\tilde{x},\tilde{y})$, it is always possible, by relabeling the
variables, to rewrite integral in terms of $(x,y)$. Consequently the
invariance condition (\ref{eq:sup_inv_energy}) can be expressed in terms of the free energy density, $f$ as
\begin{equation}
  f(\rho,\nabla\rho,\nabla\nabla\rho)=f(\rho,\mathbf{R}_\theta\nabla\rho,\mathbf{R}_\theta\nabla\nabla\rho\mathbf{R}_\theta^T).
\label{eq:sup_inv_cond}
\end{equation}
  
If we assume $f$ to have a polynomial form (in $\rho$, $\rho_x$,
$\rho_y$, $\rho_{xx}$, $\rho_{xy}$ and $\rho_{yy}$), then equation
(\ref{eq:sup_inv_cond}) imposes constraints on its
coefficients. These constraints take the form of linear equations
whose solution(s) gives us the required form(s) for the rotationally
invariant free energy. In this work, we have considered the cases when
$f$ is a polynomial of degree two, three and four, respectively. For
each of these cases, we used Mathematica to enforce condition
(\ref{eq:sup_inv_cond}) and to solve the resulting system of linear
equations.

We summarize the essential steps for the case of quadratic order and
leave it up to the reader to verify our observations for the cubic and
quartic orders.
\begin{enumerate}
  \item Consider a general quadratic polynomial in $\rho$, $\rho_x$,
    $\rho_y$, $\rho_{xx}$, $\rho_{xy}$ and $\rho_{yy}$ for $f$:
    \begin{multline}
      f=
  a_1 \rho_{xx}^2+
  a_2 \rho_{xx} \rho_{xy}+
  a_3 \rho_{xx}\rho_{yy}+
  a_4 \rho_{x} \rho_{xx}+
  a_5 \rho_{xx} \rho_{y}+
  a_6 \rho  \rho_{xx}+
  a_7 \rho_{xy}^2+
  a_8 \rho_{xy} \rho_{yy}+
  a_9 \rho_{x} \rho_{xy}+
  a_{10} \rho_{xy} \rho_{y}+\\
  a_{11} \rho  \rho_{xy}+
  a_{12} \rho_{yy}^2+
  a_{13} \rho_{x} \rho_{yy}+
  a_{14}\rho_{y} \rho_{yy}+
  a_{15}\rho  \rho_{yy}+
  a_{16} \rho_{x}^2+
  a_{17} \rho_{x} \rho_{y}+
  a_{18} \rho  \rho_{x}+ 
  a_{19} \rho_{y}^2+
  a_{20} \rho  \rho_{y}+
  a_{21} \rho^2,
\end{multline}
where the coefficients $a_1,\cdots a_{21}$ are to be determined.
\item 
The invariance condition (\ref{eq:sup_inv_cond}) implies:
$a_7=2a_1-a_3$, $a_{12}=a_{1}$, $a_{15}=a_{6}$, $a_{19}=a_{16}$ and
all $a_i$ are zero, \emph{except} for $i=1,3,6,16$ and $21$. Thus, we
obtain
$$f=a_1 \Big(\rho_{xx}^2+\rho_{yy}^2+2\rho_{xy}\Big)+a_3
\Big(\rho_{xx}\rho_{yy}-\rho_{xy}^2\Big) +
a_6\rho(\rho_{xx}+\rho_{yy})+a_{16}(\rho_x^2+\rho_y^2)+a_{21}\rho^2,$$
which in a coordinate-independent form may be written as
\begin{equation}
f=a_1
(\Delta\rho)^2+b\det(\nabla\nabla\rho)+a_6\rho\Delta\rho+a_{16}|\nabla\rho|^2+a_{21}\rho^2,
\label{eq:sup_quad}
\end{equation}
where $b=a_3-2a_1$. On a sphere these derivatives must be interpreted
by their covariant counterparts.

At quadratic order we find no chiral terms in (\ref{eq:sup_quad}) and
for $a_1=1/2$, $b=0$, $a_6=k_0^2$, $a_{16}=0$ and $a_{21}=(k_0^2+r)/2$
we recover the quadratic part of the Landau-Brazovskii energy density (1).
\end{enumerate}

Repeating the procedure at cubic order, we find only one chiral term:
$$f_3:=\nabla\rho\cdot(\nabla\nabla\rho)\cdot(\mathbf{k}\times\nabla\rho),$$
while at quartic order, we find two:
$$f_4^a:=\rho\nabla\rho\cdot(\nabla\nabla\rho)\cdot(\mathbf{k}\times\nabla\rho)$$
and
$$f_4^b:=\nabla\rho\cdot(\nabla\nabla\rho)^2\cdot(\mathbf{k}\times\nabla\rho).$$
On the surface of a sphere the partial derivatives are replaced by
their covariant counterparts and $\mathbf{k}$ is replaced by the normal
to the sphere $\mathbf{n}$.

We now show that,
\begin{equation}
  I_3 = \int_{S^2} f_3\;dS=0.
  \label{eq:sup_trivial1}
\end{equation}
Consider
\begin{equation}
  I_3=\int_{S^2} \nabla\rho\cdot \nabla\nabla\rho\cdot(\mathbf{n}\times\nabla\rho)\;dS.
\end{equation}
which explicitly in terms of coordinates can be written as
$$I_3=\int_{S^2}(\nabla^\alpha \rho) (\nabla_\alpha\nabla_\beta\rho) (\varepsilon^{\beta\nu}\nabla_\nu\rho)\;dS.$$
Integrating by parts,
\begin{equation}
  I_3=\int_{S^2}\nabla^\alpha\Big[\rho \varepsilon^{\beta\nu}\nabla_\nu\rho (\nabla_\alpha\nabla_\beta\rho)\Big]-\rho\nabla^\alpha\Big[\varepsilon^{\beta\nu}\nabla_\nu\rho (\nabla_\alpha\nabla_\beta\rho)\Big]\;dS.
\end{equation}
Since a sphere has no boundary, the first term is zero by divergence
theorem. Expanding the second term and using the definition of the
Laplace-Beltrami, $\nabla^\alpha\nabla_\alpha=:\Delta$,
$$I_3=-\int_{S^2}\rho
\Delta(\nabla_\beta\rho)(\varepsilon^{\beta\nu}\nabla_\nu\rho)+\rho(\nabla_\alpha\nabla_\beta\rho)\varepsilon^{\beta\nu}\nabla^\alpha\nabla_\nu\rho\;dS.$$
The second term involves the product of the anti-symmetric
$\varepsilon^{\beta\nu}$ and the symmetric
$\nabla_\alpha(\nabla_\beta\rho)\nabla^\alpha(\nabla_\nu\rho)$ and is
therefore zero. Using the following identity on the unit-sphere,
$$\Delta(\nabla_\beta)\rho=\nabla_\beta(\Delta\rho)+\nabla_\beta\rho,$$
we obtain,
$$I_3=-\int_{S^2}\rho
\varepsilon^{\beta\nu}\nabla_\beta(\Delta\rho)\nabla_\nu\rho + \rho
\varepsilon^{\beta\nu}\nabla_\nu\rho\nabla_\beta\rho\;dS.$$ Again, the second
term in this expression is zero as it involves the product of symmetric
and antisymmetric tensors. Applying divergence theorem on the
first term,
$$I_3=\int_{S^2}(\Delta\rho)\varepsilon^{\beta\nu}\Big[\nabla_\beta\rho\nabla_\nu\rho+\rho\nabla_\beta\nabla_\nu\rho\Big]\;dS=0,$$
where the last equality follows from the observation that the
integrand is the product of symmetric and anti-symmetric tensors, thus
establishing (\ref{eq:sup_trivial1}).

Similarly, it can be shown that one of the chiral quartic terms,
$$I_4^a=\int_{S^2}f_4^a\;dS=\int_{S^2}\rho \nabla\rho\cdot \nabla\nabla\rho\cdot(\mathbf{n}\times\nabla\rho)\;dS = 0.$$

The only non-trivial chiral term that we find is the quartic term:
\begin{equation}
  I_4^b=\int_{S^2}f_4^b\;dS = \int_{S^2} \nabla\rho\cdot (\nabla\nabla\rho)^2\cdot(\mathbf{n}\times\nabla\rho)\;dS.
  \label{eq:sup_quartic_chiral}
\end{equation}
\section{S2: Projecting Euler-Lagrange Equations}
\label{sec:EL-project}
If we expand the density modulation $\rho$ in terms of spherical
harmonics ($Y_l^m$, $l=1,2,3,\cdots;-l\leq m \leq l$) then close to the
melting transition the behavior of the system is dominated by modes
`$l$' selected by the parameter $k_0$. Choosing $k_0^2=l(l+1)$ ensures
that the transition occurs at $r=0$. With this choice, we restrict the
density expansion in terms of modes $l$:
\begin{equation}
  \rho = \sum_{m=-l}^l c_{m}Y_{lm}.
  \label{eq:sup_density_expansion}
\end{equation}
where $c_m$ are the modal amplitudes. For real densities
$c^*_{m}=(-1)^mc_{-m}$.

Let us now consider the Landau-Brazovskii energy,
\begin{equation}
F_{LB} = \int_{S^2} \Big[(\Delta+k_0^2)\rho\Big]^2+\frac{r}{2}\rho^2+\frac{u}{3}\rho^3+\frac{w}{4}\rho^4\;da,
  \label{eq:sup_LB energy}
\end{equation}
whose Euler-Lagrange equation is given by
\begin{equation}
 \frac{\delta{F_{LB}(\rho)}}{\delta\rho}:=(\Delta+k_0^2)^2\rho + r\rho + u\rho^2+w\rho^3 = 0.
\label{eq:sup_PDE}
\end{equation}

Assuming $k_o^2=l(l+1)$, we project the Euler-Lagrange equation onto
the subspace of spherical harmonics of order $l$. One way to achieve this
is by plugging in (\ref{eq:sup_density_expansion}) into (\ref{eq:sup_PDE})
and integrating with respect to $Y_l^m$,
\begin{equation}
  \langle Y_l^m, \frac{\delta{F_{LB}(\rho)}}{\delta\rho}\rangle=G_m(c_{-l},c_{-l+1},\cdots,c_0,\cdots,c_{l-1},c_{l};r,u,w)=0,\;-l\leq m\leq l.
  \label{eq:sup_EL}
\end{equation}
Although the explicit form for $G_m$ may be obtained, brute-force, by
symbolic integration in Mathematica or by using the Wigner $3j$
symbols to compute products of spherical harmonics, in practice, we
found these methods to extremely time-consuming. Instead, we use an
elegant and very efficient algorithm developed by Sattinger
\cite{sattinger1978bifurcation} based on ladder operators of quantum mechanics, which
constitute the generators of the Lie algebra of $SO(3)$. This method
generates the most general polynomial in $2l+1$ variables
($c_{-l},\cdots,c_0,\cdots,c_l$) which is
$SO(3)$-equivariant. However, some coefficients of this polynomial
remain arbitrary. The generality of the polynomial means that any
$SO(3)$-symmetric theory where the order parameter has a polynomial
dependence can be written in terms of the these polynomials. To
specialize to our particular theory (with free energy (\ref{eq:sup_LB
  energy})), we will need to compute as many terms of $\langle Y_l^m,
\frac{\delta{F_{LB}(\rho)}}{\delta\rho}\rangle$ by explicit
integrations as required to remove the arbitrariness of the
coefficients. The number of integrations required in this method is
far lower than in the ``brute-force'' method, mentioned earlier. In
the $l=6$ case, four integral computations were required.
\subsection{Case Study: $l=6$}
Once the projected Euler-Lagrange equations (\ref{eq:sup_EL}) are
constructed, we analyze the system for icosahedrally symmetric
equilibria by plugging in an ansatz. For the purposes of illustration,
we present the details of our analysis when $l=6$. Since
the cubic term in the equations are quite complicated we do not show
them in the discussion below. However, for the convenience of a
curious reader, these are summarized at the end of this supplementary
material in section (S8). We have,
\begin{subequations}
\begin{equation}
G_6:=  c_6 r-\frac{10u}{323} \sqrt{\frac{13}{231 \pi }} \Big(-42 c_3^2+14 \sqrt{30} c_2 c_4-7 \sqrt{66} c_1 c_5+2 \sqrt{231} c_0 c_6\Big)+\cdots  =0,
  \label{eq: 1}
\end{equation}
\begin{equation}
G_5 := c_5 r+\frac{10u}{323} \sqrt{\frac{13}{77 \pi }} \Big(14 c_2 c_3-21 \sqrt{3} c_1 c_4+5 \sqrt{77} c_0 c_5-7 \sqrt{22} c_{-1} c_6\Big)+\cdots =0,
    \label{eq: 2}
\end{equation}
\begin{equation}
G_4 := c_4 r-\frac{5u \sqrt{\frac{26}{21 \pi }}}{3553} \Big(-84 \sqrt{3} c_2^2+35 \sqrt{30} c_1 c_3+8 \sqrt{42} c_0 c_4-63 \sqrt{22} c_{-1} c_5+28 \sqrt{165} c_{-2} c_6\Big) +\cdots =0,
    \label{eq: 3}
\end{equation}
\begin{equation}
G_3 := c_3 r+\frac{u\sqrt{\frac{130}{77 \pi }}}{3553} \Big(42 \sqrt{165} c_1 c_2-43 \sqrt{770} c_0 c_3+7 \Big(25 \sqrt{22} c_{-1} c_4+22 \sqrt{10} c_{-2} c_5-44 \sqrt{30} c_{-3} c_6\Big)\Big)+\cdots  = 0,
    \label{eq: 4}
\end{equation}
\begin{multline}
G_2:= c_2 r-\frac{10u \sqrt{\frac{13}{231 \pi }}}{3553} \Big(-42 \sqrt{55} c_1^2+22 \sqrt{231} c_0 c_2+7 \Big(9 \sqrt{22} c_{-1} c_3\\-12 \sqrt{66} c_{-2} c_4+22 \sqrt{3} c_{-3} c_5+22 \sqrt{30} c_{-4}
   c_6\Big)\Big)+\cdots = 0,
    \label{eq: 5}
\end{multline}
\begin{multline}
G_1:= c_1 r+\frac{u\sqrt{\frac{130}{231 \pi }}}{3553} \Big(20 \sqrt{2310} c_0 c_1-7 \Big(60 \sqrt{22} c_{-1} c_2-18 \sqrt{55} c_{-2} c_3\\-25 \sqrt{66} c_{-3} c_4+99 \sqrt{10} c_{-4} c_5+22 \sqrt{165} c_{-5}
   c_6\Big)\Big)+\cdots = 0,
    \label{eq: 6}
\end{multline}
\begin{equation}
G_0:= c_0 r+\frac{10u \sqrt{\frac{13}{\pi }}}{3553} \Big(20 c_0^2-20 c_{-1} c_1-22 c_{-2} c_2+43 c_{-3} c_3-8 c_{-4} c_4-55 c_{-5} c_5-22 c_{-6} c_6\Big)+\cdots = 0.
    \label{eq: 7}
\end{equation}
\end{subequations}
Equations for $m<0$ can be obtained by replacing $c_m$ by $c_{-m}$.

Plugging the icosahedral ansatz:
\begin{equation}
\rho_{I}= \xi\mathcal{Y}_h(6) = \xi\Big(Y_{0,0}+\sqrt{\frac{7}{11}}Y_{6,5}-\sqrt{\frac{7}{11}}Y_{6,-5}\Big),
  \label{eq:ansatz}
\end{equation}
\emph{i.e.,}
\begin{equation}
(c_{-6},\cdots,c_0,\cdots,c_6) = (0,-\xi\sqrt{\frac{7}{11}},0,0,0,0,\xi,0,0,0,0,\xi\sqrt{\frac{7}{11}},0),
\end{equation}
in terms of the unknown amplitude $\xi$ into (\ref{eq: 1}-\ref{eq: 7})
(and the $m<0$ counterparts), we find that (\ref{eq: 1}),(\ref{eq:
  3})-(\ref{eq: 6}) are trivially zeros. Equations (\ref{eq: 2}) and
(\ref{eq: 7}) both reduce to a cubic equation in $\xi$:
\begin{equation}
r\xi+\frac{50u\sqrt{13}}{323\sqrt{\pi}}\xi^2+\frac{2145 w}{1564\pi}\xi^3=0.
\label{eq:sup_char_eq1}
\end{equation}
The equations $G_m$ for $m<0$ do not add any new equations.

The plot showing various ``branches'' of solutions to
(\ref{eq:sup_char_eq1}) (for $u=-10$, $w=10$) as a function of $r$ is
shown in Fig. (\ref{fig:sup_curves}). The stability of these branches,
determined by computing the eigen values of the hessian (obtained by
differentiating (\ref{eq: 1}-\ref{eq: 7} and the $m<0$ counterparts),
indicate a clear first order-transition from a liquid state ($\xi=0$)
to an icosahedral modulated state (shown in solid blue in
Fig. (\ref{fig:sup_curves}). The red curves are unstable. The
meta-stable coexistence states are shown as blue dashed curves.
\begin{figure}
  \begin{center}
  \includegraphics[width=3in]{l6solution.pdf}
  \end{center}
  \caption{Plot showing various branches of Eq. (\ref{eq:sup_char_eq})
    for $l=6$. Solid Blue curves are states with the lowest energy (at
    a particular $r$), dashed blue are local minimizers of energy and
    red dashed curve represent unstable states.}
    \label{fig:sup_curves}
\end{figure}

\section{S3: Chiral Term}
In this section we show that the if the density $\rho$ is written in
terms of a single representation expansion, \emph{i.e,}
$$\rho=\sum_{m=-l}^lc_mY_l^m,$$ then the chiral term of the free energy
\begin{equation}
  I = \int_{S^2}\nabla\rho\cdot(\nabla\nabla\rho)^2\cdot(\mathbf{n}\times\nabla\rho)\;dS=0.
\end{equation}

To see this, rewrite $I$ in its coordinate representation,
$$I=\int_{S^2}(\nabla^\alpha\nabla_\beta\rho)(\nabla^\beta\rho)(\nabla_\alpha\nabla_\nu\rho)(\varepsilon^{\nu\gamma}\nabla_{\gamma}\rho)\;dS.$$
Using product rule on the first two product terms of the integrand,
this can written as,
$$I=\frac{1}{2}\int_{S^2}\nabla^\alpha(|\nabla\rho|^2)(\nabla_\alpha\nabla_\nu\varepsilon^{\nu\gamma}\nabla_\gamma\rho)\;dS.$$
Now using divergence theorem, we obtain
$$2I = \int_{S^2}\nabla^\alpha\Big[|\nabla\rho|^2(\nabla_\alpha\nabla_\nu\varepsilon^{\nu\gamma}\nabla_\gamma\rho)\Big]-|\nabla\rho|^2\nabla^\alpha[(\nabla_\alpha\nabla_\nu\rho)\varepsilon^{\nu\gamma}\nabla_\gamma\rho]\;dS.$$
Dropping the first surface term (since $S^2$ has no boundary) and expanding the gradient in the second term, we have
$$2I =
-\int_{S^2}|\nabla\rho|^2\Big[\Delta(\nabla_\nu\rho)\varepsilon^{\nu\gamma}\nabla_\gamma\rho
  +
  \varepsilon^{\nu\gamma}(\nabla_\alpha\nabla_\nu\rho)(\nabla^\alpha\nabla_\gamma\rho)\Big]\;dS.$$
The second term, being a product of a symmetric and an anti-symmetric
tensor, evaluates to zero. Using the identity, $\Delta(\nabla_\nu\rho)
= \nabla_\nu(\Delta\rho) + \nabla_\nu\rho$, for unit sphere, we can
rewrite the first term in previous integral as,
\begin{equation}
  I=-\frac{1}{2}\int_{S^2}|\nabla\rho|^2 \nabla(\Delta\rho)\times\nabla\rho\;dS,
  \label{eq:sup_chiral}
\end{equation}
having eliminated the term involving the product
$\varepsilon^{\nu\gamma}\nabla_\nu\rho\nabla_\gamma\rho$ as zero. It
is now clear that if $\rho=\sum_{m=-l}^lc_mY_l^m$ then
$\Delta\rho=-l(l+1)\rho$. Consequently, due to the cross product term
in the integrand, the integral (\ref{eq:sup_chiral}) evaluates to
zero.

We wish to point out that $I$ is zero only when $\rho$ is expressed in
terms of a single representation. However, in coupled representation, for
instance,
$$\rho = \xi\mathcal{Y}_h(16)+\zeta\mathcal{Y}_h(15),$$ where
$\mathcal{Y}_h(16)$ and $\mathcal{Y}_h(15)$ are icosahedrally
symmetric spherical harmonics for $l=16$ and $l=15$, respectively,
direct integration shows that $I$ is not zero. 

\section{S4: Density Modulation for $l=6, 10$ and $12$}
In Fig. (\ref{fig:l=6,10,12}) we compare the density modulations for
$l=6$, $l=10$ and $l=12$ to Caspar-Klug T-number constructions for
viral capsids corresponding to $T=1$, $T=3$ and $T=4$, respectively.

\begin{figure}[h!]
  \begin{center}
  \includegraphics[width=3in]{OrderPar}
  \end{center}
   \caption{(top) Density modulation profiles obtained from Landau-Brazovskii theory for $l$=6, 10 and 12. (bottom) T=1, T=3, and T=4 Caspar-Klug icosahedra obtained from Monte Carlo simulations of N=12, 32, and 42 interacting two-state ``capsomer particles" (from ref.\cite{Zandi}). The first state (smaller red disks) represents pentameric capsomers composed of five chiral subunits while the second state (larger blue disks) represents hexameric capsomers composed of six chiral subunits.}
    \label{fig:l=6,10,12}
\end{figure}

Similar to the case of $l=6$ discussed in section (S2), we find that
for segments corresponding to $l=10$ and $l=12$ the order parameter
exhibits a first order solidification transition to icosahedral states
shown above. The bifurcation diagrams for these two cases are
schematically similar to Fig. (\ref{fig:sup_curves}).

\section{S5: Instability of the $l=15$ Mode}
In this section we show that the $l=15$ mode is locally unstable by
showing that the second variation of the free energy is negative for
certain perturbations of the equilibrium state.

Let us begin by noting that the icosahedral equilbrium state for
$l=15$ may be obtained by following a procedure detailed in section
(S2). It is also possible to obtain this by directly integrating the
free energy (\ref{eq:sup_LB energy}) with the icosahedral ansatz:
\begin{equation}
  \rho_{I} = \zeta\mathcal{Y}_h(15)=\zeta i \Big[ (Y_{15}^5+Y_{15}^{-5}) - \frac{\sqrt{7590}}{115}(Y^{10}_{15}-Y^{-10}_{15})- \frac{\sqrt{3338335}}{3335}(Y^{15}_{15}+Y^{-15}_{15})\Big]
  \label{eq:sup_y15}
\end{equation}
and setting the derivative of the resulting expression with respect to
$\zeta$ to zero. Either procedures lead to the following cubic
equation for the unknown amplitude $\zeta$:
\begin{equation}
  \zeta\Big(584803025179 \pi r + 1449757237500 w \zeta^2\Big)=0,
  \label{eq:sup_char_eq}
\end{equation}
which can be readily solved yielding,
\begin{equation}
  \zeta = \zeta^* := \pm 1.126\frac{\sqrt{-r}}{\sqrt{w}},\; (r< 0).
  \label{eq:sup_l15_sol}
\end{equation}
Here, we have discarded the trivial solution $\zeta=0$. Note
that due to the anti-symmetry of the odd spherical hamonics $l=15$,
the cubic terms of the free energy (\emph{i.e.,} terms proportional to
$u$) integrate to zero and therefore make no contributions
to (\ref{eq:sup_char_eq}).

The stability may be determined by computing the eigen values of the
hessian obtained by differentiating the projected Euler-Lagrange
equations using the procedure discussed earlier. However, here, we
provide an alternate method to convince the reader that the $l=15$
mode is indeed unstable.

We perturb about this icosahedral state:
$$\rho=\zeta^* \mathcal{Y}_{h}(15)+\hat{\rho},$$ where
$\hat\rho:=\sum_{m=-15}^{15}c_mY^m_{15}$ is the perturbation in the
space of $l=15$ spherical harmonics. The perturbation in energy up
to quadratic order is given by
\begin{equation}
\delta F =\int_{S^2}\frac{r}{2}\hat\rho^2
+\frac{3}{2}w(\zeta^*)^2\mathcal{Y}_{15}^2\hat\rho^2\;dS.
\label{eq:sup_second_var}
\end{equation}
The term involving $u$ integrates to zero by its anti-symmetry.

We claim that for perturbation $\hat{\rho}=Y_{15}^0$ (along
axisymmetric spherical harmonic for $l=15$), the variation in the free
energy is negative and consequently destabilizes the system. This may
be verified by direct integration of (\ref{eq:sup_second_var}) by
substitution for $\zeta^*$ from (\ref{eq:sup_l15_sol}) along with
$\hat{\rho}=Y_{15}^0$. Using Mathematica, we obtain
$$\delta F = 0.003261 r < 0\text{ (since }r<0),$$ thus establishing
the instability of the $l=15$ mode.

\section{S6: The $15\times 16$ Superposition State}
The $15\times 16$ superposition state is constructed by taking multiple
irreps of the icosahedral group (c.f. Eq. (4) of the main article):
$$\rho(\zeta,\xi)=\zeta\mathcal{Y}_h(15) + \xi \mathcal{Y}_h(16),$$
where $\zeta$ and $\xi$ are the amplitudes of the icosahedrally
symmetric $l=15$ and $16$ spherical harmonics -- $\mathcal{Y}_h(15)$
and $\mathcal{Y}_h(16)$, respectively. The solutions to the coupled
cubic system Eq. (5) (of the main article) for $k_0=16$ has been
plotted in Fig. (\ref{fig:xivszeta}). We see a weakly-first order
transition to an icosahedrally symmetric density profile. For
$\chi=0$, there are two degenerate solutions marked as $D$ and
$L$. The corresponding density modulations are shown in top part of
Fig. (\ref{fig:T7}). In the bottom part of this figure, we compare the
density profiles with the results of a two-state MC simulation
\cite{Zandi} of a $T=7$ chiral CK capsid composed of 72 caspsomers.
\begin{figure}[h!]
  \begin{center}
    \begin{tabular}{cc}
      \includegraphics[width=3in]{Figure2_a} &
      \includegraphics[width=3in]{Figure2_b}
      \end{tabular}
  \end{center}
 \caption{Solution branches of Eq. (5) for $k_0=16$, $u$ = -10,
   $w=10$, and $\chi=0$. Only solution branches with stable sections
   are shown. (left) Coefficient $\zeta$ of $\mathcal{Y}_h(15)$ as a
   function of the control parameter $r$. Blue: stable sections. Red:
   unstable sections. The two branches marked $D$ and $L$ are
   degenerate. (right) Coefficient $\xi$ of $\mathcal{Y}_h(16)$ as a
   function of $\zeta$ for varying $r$.}
    \label{fig:xivszeta}
\end{figure}

\begin{figure}[h!]
  \begin{center}
  \includegraphics[width=3in]{15x16}
  \end{center}
  \caption{(top) The $D$ and $L$ isomers of $15\times16$ for
    $k_0=16$. (bottom) T=7 chiral Caspar-Klug isomers obtained from
    Monte Carlo simulations of N=72 interacting two-state capsomer
    particles. From ref.\cite{Zandi}. (top, right) Alternative subunit
    interpretation as a chiral T=1 CK capsid (subunits outlined in
    black, pentamers in red.)}
    \label{fig:T7}
\end{figure}

\section{S7: Icosahedrally Symmetric Energy}
Recall (from the discussion in the letter) that states with symmetries $C_5$, $C_3$ and $C_2$ that break
the icosahedral symmetry of the $15\times 16$ state (cf. Fig. 4A)
are associated with a three dimensional irreducible representation of
the icosahedral group. More specifically, it is the irreducible
representation referred to as `3' in \cite{Hoyle}, whose generators are
given by:
\begin{equation}
  \mathbf{M}_5 = \left(\begin{array}{ccc}
    -\tau &-1/\tau & 1\\1/\tau & 1 & \tau\\-1 &\tau &-1/\tau
  \end{array}
  \right),\;   \mathbf{M}_3 = \left(\begin{array}{ccc}
    -1 & 0 & 0\\0 & 1 & 0\\0 & 0 & -1
  \end{array}
  \right),
\end{equation}
where $\tau=(1+\sqrt{5})/2$. Let us denote the icosahedral group
generated by these matrices by $\mathbb{I}_3$.

In this representation, we can associate a 3d-vector order parameter,
$(\eta_1,\eta_2,\eta_3)$, that measures the symmetry-breaking from the
icosahedral state. An orthogonal basis for this three dimensional
vector space may chosen such that $(1,1,1)$ corresponds to the $C_3$
state and $(0,0,1)$ corresponds to the $C_2$ state. The general form
of the free energy for the order parameter $(\eta_1,\eta_2,\eta_3)$
(that is equivalent to Eq. (6) of the main text) has been derived by
Hoyle in \cite{Hoyle}. For readers unfamiliar with equivariant
bifurcation theory, we provide an alternate derivation.

To construct the free energy (6), we enforce the invariance of the energy under
the three dimensional representation of the icosahedral group
$\mathbb{I}_3$ using a procedure similar to the one discussed in
section (S1). We outline the steps for convenience,
\begin{enumerate}
\item Expand the energy as a general polynomial
  $p_n(\eta_1,\eta_2,\eta_3)$ of order $n$ in $\eta_1$, $\eta_2$ and
  $\eta_3$, where $n=2,3,4,5$, and $6$. We wish to determine the
  coefficients of the polynomial.
\item Enforce the invariance of the energy under $\mathbb{I}_3$ via
  \begin{equation}
    p_n\Big(\mathbf{M}(\eta_1,\eta_2,\eta_3)\Big) = p_n(\eta_1,\eta_2,\eta_3),\text{ for all }\mathbf{M}\in\mathbb{I}_3
    \label{eq:sup_inv_cond2}
  \end{equation}
\item Comparing the coefficients of various terms in
  (\ref{eq:sup_inv_cond2}) leads to a system of linear equations for the coefficients of $p_n$.
\item The solutions of this system provide the form of the required
  invariant energy.
\end{enumerate}
It is quite straightforward to automate the procedure using computer
algebra. We provide the details for the sixth order term in the
energy. Consider the sixth order polynomial in $\eta_1$, $\eta_2$ and
$\eta_3$:
  \begin{multline}
    p_6 = a_1 \eta _1^6+a_2 \eta _2 \eta _1^5+a_3 \eta _3 \eta
    _1^5+a_4 \eta _2^2 \eta _1^4+a_5 \eta _2 \eta _3 \eta _1^4+a_6
    \eta _3^2 \eta _1^4+a_7 \eta _2^3 \eta _1^3+a_8 \eta _2^2 \eta _3
    \eta _1^3+a_9 \eta _2 \eta _3^2 \eta _1^3+\\ a_{10} \eta _3^3 \eta
    _1^3+a_{11} \eta _2^4 \eta _1^2+a_{12} \eta _2^3 \eta _3 \eta
    _1^2+a_{13} \eta _2^2 \eta _3^2 \eta _1^2+a_{14} \eta _2 \eta _3^3
    \eta _1^2+a_{15} \eta _3^4 \eta _1^2+a_{16} \eta _2^5 \eta
    _1+a_{17} \eta _2^4 \eta _3 \eta _1+\\a_{18} \eta _2^3 \eta _3^2
    \eta _1+a_{19} \eta _2^2 \eta _3^3 \eta _1+a_{20} \eta _2 \eta
    _3^4 \eta _1+a_{21} \eta _3^5 \eta _1+a_{22} \eta _2^6+a_{23} \eta
    _2^5 \eta _3+a_{24} \eta _2^4 \eta _3^2+a_{25} \eta _2^3 \eta
    _3^3+a_{26} \eta _2^2 \eta _3^4+a_{27} \eta _2 \eta _3^5+a_{28}
    \eta _3^6
  \end{multline}
  with $a_1, a_2 \cdots a_{28}$ as the coefficients. The invariance
  condition (\ref{eq:sup_inv_cond2}) can be equivalently written as a
  homogeneous linear system of equations for the coefficients. We find
  two linearly independent solutions which provide the following
  general form for the invariant energy:
\begin{equation}
  p_6 = a p_a + b p_b,
\label{eq:sup_gen_form}
\end{equation}
  where,
  \begin{multline}
    p_a = 0.108745 \eta _1^6+0.458216 \eta _2^2 \eta _1^4+0.275823 \eta _3^2 \eta _1^4+0.275823 \eta _2^4 \eta _1^2+0.458216 \eta _3^4 \eta _1^2+\\0.326196 \eta _2^2 \eta _3^2 \eta _1^2+0.108745 \eta
    _2^6+0.108745 \eta _3^6+0.275823 \eta _2^2 \eta _3^4+0.458216 \eta _2^4 \eta _3^2,
  \end{multline}
  \begin{multline}
    p_b = -0.0089041 \eta _1^6+0.287171 \eta _2^2 \eta _1^4-0.146605 \eta _3^2 \eta _1^4-0.146605 \eta _2^4 \eta _1^2+0.287171 \eta _3^4 \eta _1^2-\\0.829386 \eta _2^2 \eta _3^2 \eta _1^2-0.0089041 \eta
    _2^6-0.0089041 \eta _3^6-0.146605 \eta _2^2 \eta _3^4+0.287171 \eta _2^4 \eta _3^2,
  \end{multline}
  where $a$ and $b$ are arbitrary.
  
  To relate the general form (\ref{eq:sup_gen_form}) to (6), we
  reparametrize $a$ and $b$ by choosing
  $$a=1.48163 W + 0.240062 \Delta,$$
  $$b=-0.62299W -1.25361 \Delta,$$
  resulting in,
  $$p_6 = \frac{W}{6}(\eta_1^2+\eta_2^2+\eta_3^2)^3 + \Delta\Big[ \frac{\sqrt{5}}{2}\eta_1^2\eta_2^2\eta_3^2+\frac{\sqrt{5}}{60}(\eta_1^6+\eta_2^6+\eta_3^6)+\frac{1}{4}\Big(\eta_1^4(\eta_3^2-\eta_2^2) + \eta_2^4(\eta_1^2-\eta_3^2) + \eta_3^4(\eta_2^2-\eta_3^2) \Big)\Big],$$
  which is the desired sixth order contribution to (6).
\section{S8: Summary of Projected Euler Lagrange Equations for $l=6$}
As a reference, we provide the explicit form for the projected
Euler-Lagrange equations for $l=6$ up to cubic order.
\begin{multline}
  G_6:= r c_6-\frac{10}{323} \sqrt{\frac{13}{231 \pi }} u \Big(-42
  c_3^2+14 \sqrt{30} c_2 c_4-7 \sqrt{66} c_1 c_5+2 \sqrt{231} c_0
  c_6\Big)+\\ \frac{38766 w}{37145 \pi } \Big(\frac{14}{9}
  \sqrt{\frac{5}{11}} c_2^3-7 \sqrt{\frac{2}{11}} c_1 c_3
  c_2+\sqrt{\frac{70}{11}} c_0 c_4 c_2-\sqrt{\frac{10}{3}} c_{-1} c_5
  c_2+c_{-2} c_6 c_2+2 \sqrt{\frac{21}{11}} c_0 c_3^2+9
  \sqrt{\frac{5}{11}} c_{-2} c_4^2+5 \sqrt{\frac{22}{3}} c_{-4}
  c_5^2+35 c_{-6} c_6^2-\\ 6 \sqrt{\frac{15}{11}} c_{-1} c_3 c_4+4
  \sqrt{3} c_{-2} c_3 c_5-5 \sqrt{10} c_{-3} c_4 c_5-5 c_{-3} c_3
  c_6+15 c_{-4} c_4 c_6-35 c_{-5} c_5 c_6\Big)\\ -\frac{372099
    w}{371450 \pi } \Big(\frac{56}{9} \sqrt{\frac{5}{11}} c_2^3-28
  \sqrt{\frac{2}{11}} c_1 c_3 c_2+2 \sqrt{\frac{70}{11}} c_0 c_4
  c_2-\sqrt{\frac{10}{3}} c_{-1} c_5 c_2+8 \sqrt{\frac{21}{11}} c_0
  c_3^2+30 \sqrt{\frac{5}{11}} c_{-2} c_4^2+\\ 31 \sqrt{\frac{11}{6}}
  c_{-4} c_5^2+ 104 c_{-6} c_6^2+7 \sqrt{\frac{3}{22}} c_1^2 c_4-20
  \sqrt{\frac{15}{11}} c_{-1} c_3 c_4-\sqrt{\frac{7}{2}} c_0 c_1
  c_5+11 \sqrt{3} c_{-2} c_3 c_5-31 \sqrt{\frac{5}{2}} c_{-3} c_4
  c_5+c_{-1} c_1 c_6-\\ 11 c_{-3} c_3 c_6+42 c_{-4} c_4 c_6-104 c_{-5}
  c_5 c_6\Big)+\\ \frac{367497 w}{742900 \pi } \Big(\frac{28}{3}
  \sqrt{\frac{5}{11}} c_2^3-42 \sqrt{\frac{2}{11}} c_1 c_3 c_2+2
  \sqrt{\frac{70}{11}} c_0 c_4 c_2+12 \sqrt{\frac{21}{11}} c_0
  c_3^2+42 \sqrt{\frac{5}{11}} c_{-2} c_4^2\\ +7 \sqrt{66} c_{-4}
  c_5^2+140 c_{-6} c_6^2+7 \sqrt{\frac{6}{11}} c_1^2 c_4-28
  \sqrt{\frac{15}{11}} c_{-1} c_3 c_4-\sqrt{14} c_0 c_1 c_5+14
  \sqrt{3} c_{-2} c_3 c_5-\\ 21 \sqrt{10} c_{-3} c_4 c_5+c_0^2 c_6-14
  c_{-3} c_3 c_6+56 c_{-4} c_4 c_6-140 c_{-5} c_5 c_6\Big)=0
\end{multline}
\clearpage
\begin{multline}
  G_5:=r c_5+\frac{10}{323} \sqrt{\frac{13}{77 \pi }} u \Big(14 c_2
  c_3-21 \sqrt{3} c_1 c_4+5 \sqrt{77} c_0 c_5-7 \sqrt{22} c_{-1}
  c_6\Big)+\\ \frac{6461 \sqrt{3} w}{37145 \pi}\Big(-28
  \sqrt{\frac{5}{11}} c_3 c_1^2+\frac{14}{3} \sqrt{\frac{2}{11}} c_2^2
  c_1+20 \sqrt{\frac{7}{11}} c_0 c_4 c_1-\frac{20 c_{-1} c_5
    c_1}{\sqrt{3}}+2 \sqrt{10} c_{-2} c_6 c_1-48 \sqrt{\frac{2}{11}}
  c_{-1} c_3^2-\\ 10 \sqrt{55} c_{-3} c_4^2+54 \sqrt{\frac{5}{11}}
  c_{-3} c_4^2-70 \sqrt{3} c_{-5} c_5^2+\frac{110 c_{-5}
    c_5^2}{\sqrt{3}}+20 \sqrt{\frac{21}{11}} c_0 c_2 c_3-2 \sqrt{165}
  c_{-1} c_2 c_4-2 \sqrt{\frac{55}{3}} c_{-1} c_2 c_4+4 \sqrt{66}
  c_{-2} c_3 c_4+\\ 30 \sqrt{\frac{6}{11}} c_{-2} c_3 c_4+26 \sqrt{3}
  c_{-2} c_2 c_5-\frac{20 c_{-2} c_2 c_5}{\sqrt{3}}-36 \sqrt{3} c_{-3}
  c_3 c_5-20 \sqrt{3} c_{-4} c_4 c_5+\frac{220 c_{-4} c_4
    c_5}{\sqrt{3}}-24 c_{-3} c_2 c_6+\\ 10 \sqrt{30} c_{-4} c_3 c_6-20
  \sqrt{22} c_{-5} c_4 c_6+70 \sqrt{3} c_{-6} c_5
  c_6\Big)-\\ \frac{124033 \sqrt{3} w }{742900 \pi }\Big(-7 \sqrt{3}
  c_5 c_0^2+60 \sqrt{\frac{21}{11}} c_2 c_3 c_0-\sqrt{77} c_1 c_4
  c_0+82 \sqrt{\frac{7}{11}} c_1 c_4 c_0+\sqrt{42} c_{-1} c_6
  c_0+\frac{56}{3} \sqrt{\frac{2}{11}} c_1 c_2^2-12 \sqrt{22} c_{-1}
  c_3^2-\\ 31 \sqrt{55} c_{-3} c_4^2+180 \sqrt{\frac{5}{11}} c_{-3}
  c_4^2-208 \sqrt{3} c_{-5} c_5^2+\frac{341 c_{-5} c_5^2}{\sqrt{3}}-91
  \sqrt{\frac{5}{11}} c_1^2 c_3-2 \sqrt{\frac{55}{3}} c_{-1} c_2
  c_4-92 \sqrt{\frac{15}{11}} c_{-1} c_2 c_4+\\ 11 \sqrt{66} c_{-2}
  c_3 c_4+100 \sqrt{\frac{6}{11}} c_{-2} c_3 c_4-5 \sqrt{3} c_{-1} c_1
  c_5-\frac{20 c_{-1} c_1 c_5}{\sqrt{3}}+66 \sqrt{3} c_{-2} c_2
  c_5-\frac{20 c_{-2} c_2 c_5}{\sqrt{3}}-111 \sqrt{3} c_{-3} c_3
  c_5-\\ 71 \sqrt{3} c_{-4} c_4 c_5+\frac{682 c_{-4} c_4
    c_5}{\sqrt{3}}+2 \sqrt{10} c_{-2} c_1 c_6-66 c_{-3} c_2 c_6+31
  \sqrt{30} c_{-4} c_3 c_6-62 \sqrt{22} c_{-5} c_4 c_6+208 \sqrt{3}
  c_{-6} c_5 c_6\Big)+\\ \frac{122499 \sqrt{3} w}{1485800 \pi }
  \Big(-12 \sqrt{3} c_5 c_0^2+80 \sqrt{\frac{21}{11}} c_2 c_3 c_0-2
  \sqrt{77} c_1 c_4 c_0+124 \sqrt{\frac{7}{11}} c_1 c_4 c_0+2
  \sqrt{42} c_{-1} c_6 c_0+28 \sqrt{\frac{2}{11}} c_1 c_2^2-\\ 168
  \sqrt{\frac{2}{11}} c_{-1} c_3^2-42 \sqrt{55} c_{-3} c_4^2+252
  \sqrt{\frac{5}{11}} c_{-3} c_4^2-126 \sqrt{3} c_{-5} c_5^2-126
  \sqrt{\frac{5}{11}} c_1^2 c_3-140 \sqrt{\frac{15}{11}} c_{-1} c_2
  c_4+14 \sqrt{66} c_{-2} c_3 c_4+\\ 140 \sqrt{\frac{6}{11}} c_{-2}
  c_3 c_4-14 \sqrt{3} c_{-1} c_1 c_5+84 \sqrt{3} c_{-2} c_2 c_5-154
  \sqrt{3} c_{-3} c_3 c_5+210 \sqrt{3} c_{-4} c_4 c_5-84 c_{-3} c_2
  c_6+42 \sqrt{30} c_{-4} c_3 c_6-\\ 84 \sqrt{22} c_{-5} c_4 c_6+280
  \sqrt{3} c_{-6} c_5 c_6\Big)=0
\end{multline}
\clearpage
\begin{multline}
  G_4:=r c_4-\frac{5 \sqrt{\frac{26}{21 \pi }} u \Big(-84 \sqrt{3}
    c_2^2+35 \sqrt{30} c_1 c_3+8 \sqrt{42} c_0 c_4-63 \sqrt{22} c_{-1}
    c_5+28 \sqrt{165} c_{-2}
    c_6\Big)}{3553}+\\ \frac{19383\sqrt{\frac{2}{11}} w}{37145 \pi
  }\Big(70 \sqrt{\frac{2}{11}} c_4 c_0^2+\frac{194}{3}
  \sqrt{\frac{7}{11}} c_2^2 c_0+2 \sqrt{\frac{70}{11}} c_1 c_3 c_0-10
  \sqrt{\frac{14}{3}} c_{-1} c_5 c_0+2 \sqrt{35} c_{-2} c_6 c_0+4
  \sqrt{165} c_{-2} c_3^2-\\ 2 \sqrt{\frac{15}{11}} c_{-2}
  c_3^2+\frac{5}{3} \sqrt{22} c_{-4} c_4^2+135 \sqrt{\frac{2}{11}}
  c_{-4} c_4^2+\frac{110 c_{-6} c_5^2}{\sqrt{3}}-28
  \sqrt{\frac{15}{11}} c_1^2 c_2+\frac{56}{3} \sqrt{\frac{5}{33}}
  c_1^2 c_2-5 \sqrt{66} c_{-1} c_2 c_3- \\ 5 \sqrt{\frac{22}{3}}
  c_{-1} c_2 c_3-26 \sqrt{\frac{6}{11}} c_{-1} c_2 c_3-\frac{50}{3}
  \sqrt{22} c_{-1} c_1 c_4+70 \sqrt{\frac{2}{11}} c_{-1} c_1
  c_4+\frac{25}{3} \sqrt{22} c_{-2} c_2 c_4+90 \sqrt{\frac{2}{11}}
  c_{-2} c_2 c_4-\\ 56 \sqrt{22} c_{-3} c_3 c_4+360
  \sqrt{\frac{2}{11}} c_{-3} c_3 c_4+\frac{44}{3} \sqrt{10} c_{-2} c_1
  c_5-74 c_{-3} c_2 c_5-18 \sqrt{30} c_{-4} c_3 c_5+110
  \sqrt{\frac{10}{3}} c_{-4} c_3 c_5-\\ \frac{80}{3} \sqrt{22} c_{-5}
  c_4 c_5-6 \sqrt{30} c_{-3} c_1 c_6+18 \sqrt{10} c_{-4} c_2 c_6-10
  \sqrt{55} c_{-5} c_3 c_6+15 \sqrt{22} c_{-6} c_4
  c_6\Big)-\\ \frac{372099 w}{371450 \sqrt{22} \pi } \Big(7 \sqrt{3}
  c_6 c_{-1}^2-24 \sqrt{66} c_2 c_3 c_{-1}-5 \sqrt{\frac{22}{3}} c_2
  c_3 c_{-1}-20 \sqrt{\frac{6}{11}} c_2 c_3 c_{-1}-\frac{38}{3}
  \sqrt{22} c_1 c_4 c_{-1}-173 \sqrt{\frac{2}{11}} c_1 c_4 c_{-1}-\\ 6
  \sqrt{42} c_0 c_5 c_{-1}-5 \sqrt{\frac{21}{2}} c_0 c_5 c_{-1}-10
  \sqrt{\frac{14}{3}} c_0 c_5 c_{-1}+\frac{596}{3} \sqrt{\frac{7}{11}}
  c_0 c_2^2+3 \sqrt{165} c_{-2} c_3^2+100 \sqrt{\frac{15}{11}} c_{-2}
  c_3^2+\frac{2}{3} \sqrt{22} c_{-4} c_4^2+\\ 450 \sqrt{\frac{2}{11}}
  c_{-4} c_4^2+\frac{341 c_{-6} c_5^2}{\sqrt{3}}-91
  \sqrt{\frac{15}{11}} c_1^2 c_2+\frac{224}{3} \sqrt{\frac{5}{33}}
  c_1^2 c_2-\sqrt{\frac{385}{2}} c_0 c_1 c_3+10 \sqrt{\frac{70}{11}}
  c_0 c_1 c_3-7 \sqrt{22} c_0^2 c_4+\\ 287 \sqrt{\frac{2}{11}} c_0^2
  c_4+\frac{178}{3} \sqrt{22} c_{-2} c_2 c_4-160 \sqrt{\frac{2}{11}}
  c_{-2} c_2 c_4-122 \sqrt{22} c_{-3} c_3 c_4-111 \sqrt{\frac{11}{2}}
  c_{-3} c_3 c_4+1200 \sqrt{\frac{2}{11}} c_{-3} c_3
  c_4+\\ \frac{149}{3} \sqrt{10} c_{-2} c_1 c_5-221 c_{-3} c_2 c_5-60
  \sqrt{30} c_{-4} c_3 c_5+341 \sqrt{\frac{10}{3}} c_{-4} c_3
  c_5-\frac{128}{3} \sqrt{22} c_{-5} c_4 c_5-71 \sqrt{\frac{11}{2}}
  c_{-5} c_4 c_5+ \\ 4 \sqrt{35} c_{-2} c_0 c_6-20 \sqrt{30} c_{-3}
  c_1 c_6+60 \sqrt{10} c_{-4} c_2 c_6-31 \sqrt{55} c_{-5} c_3 c_6+42
  \sqrt{22} c_{-6} c_4 c_6\Big)+\\ \frac{367497 w}{742900 \sqrt{22}
    \pi } \Big(14 \sqrt{3} c_6 c_{-1}^2-406 \sqrt{\frac{6}{11}} c_2
  c_3 c_{-1}-14 \sqrt{22} c_1 c_4 c_{-1}-266 \sqrt{\frac{2}{11}} c_1
  c_4 c_{-1}-17 \sqrt{42} c_0 c_5 c_{-1}+\\ 268 \sqrt{\frac{7}{11}}
  c_0 c_2^2+14 \sqrt{165} c_{-2} c_3^2+28 \sqrt{\frac{15}{11}} c_{-2}
  c_3^2+630 \sqrt{\frac{2}{11}} c_{-4} c_4^2+154 \sqrt{3} c_{-6}
  c_5^2-126 \sqrt{\frac{15}{11}} c_1^2 c_2+112 \sqrt{\frac{5}{33}}
  c_1^2 c_2-\\ \sqrt{770} c_0 c_1 c_3+16 \sqrt{\frac{70}{11}} c_0 c_1
  c_3-13 \sqrt{22} c_0^2 c_4+434 \sqrt{\frac{2}{11}} c_0^2 c_4+84
  \sqrt{22} c_{-2} c_2 c_4-280 \sqrt{\frac{2}{11}} c_{-2} c_2
  c_4-\\ 245 \sqrt{22} c_{-3} c_3 c_4+1680 \sqrt{\frac{2}{11}} c_{-3}
  c_3 c_4+70 \sqrt{10} c_{-2} c_1 c_5-294 c_{-3} c_2 c_5+70 \sqrt{30}
  c_{-4} c_3 c_5-105 \sqrt{22} c_{-5} c_4 c_5+\\ 4 \sqrt{35} c_{-2}
  c_0 c_6-28 \sqrt{30} c_{-3} c_1 c_6+84 \sqrt{10} c_{-4} c_2 c_6-42
  \sqrt{55} c_{-5} c_3 c_6+56 \sqrt{22} c_{-6} c_4 c_6\Big) =0
\end{multline}
\clearpage
\begin{multline}
  G_3:=r c_3+\frac{\sqrt{\frac{130}{77 \pi }u}}{3553}\Big(42
  \sqrt{165} c_1 c_2-43 \sqrt{770} c_0 c_3+7 \Big(25 \sqrt{22} c_{-1}
  c_4+22 \sqrt{10} c_{-2} c_5-44 \sqrt{30} c_{-3}
  c_6\Big)\Big)+\\ \frac{6461 \sqrt{\frac{6}{5}} w}{37145 \pi}
  \Big(-\frac{280}{11} \sqrt{3} c_1^3+\frac{560 c_1^3}{33
    \sqrt{3}}+\frac{140}{11} \sqrt{35} c_0 c_2 c_1-\frac{298}{11}
  \sqrt{30} c_{-1} c_3 c_1+\frac{298}{11} \sqrt{10} c_{-2} c_4 c_1+8
  \sqrt{55} c_{-3} c_5 c_1-\\ 184 \sqrt{\frac{5}{11}} c_{-3} c_5
  c_1+90 \sqrt{\frac{2}{11}} c_{-4} c_6 c_1-\frac{596}{11} \sqrt{3}
  c_{-1} c_2^2+\frac{1358 c_{-1} c_2^2}{11 \sqrt{3}}-\frac{130}{11}
  \sqrt{30} c_{-3} c_3^2-25 \sqrt{22} c_{-5} c_4^2+135
  \sqrt{\frac{2}{11}} c_{-5} c_4^2+\\ \frac{84}{11} \sqrt{30} c_0^2
  c_3+\frac{235}{11} \sqrt{30} c_{-2} c_2 c_3-\frac{10}{11} \sqrt{42}
  c_{-1} c_0 c_4-\frac{1260}{11} c_{-3} c_2 c_4+\frac{256}{11}
  \sqrt{30} c_{-4} c_3 c_4-70 \sqrt{\frac{2}{11}} c_{-1}^2
  c_5+4\sqrt{\frac{770}{3}} c_{-2} c_0 c_5+\\ 2 \sqrt{\frac{210}{11}}
  c_{-2} c_0 c_5-20 \sqrt{\frac{70}{33}} c_{-2} c_0 c_5+74
  \sqrt{\frac{15}{11}} c_{-4} c_2 c_5-18 \sqrt{30} c_{-5} c_3 c_5+50
  \sqrt{3} c_{-6} c_4 c_5+14 \sqrt{\frac{15}{11}} c_{-2} c_{-1}
  c_6-\\ 12 \sqrt{\frac{70}{11}} c_{-3} c_0 c_6-12 \sqrt{10} c_{-5}
  c_2 c_6+5 \sqrt{30} c_{-6} c_3 c_6\Big)-\\ \frac{124033
    \sqrt{\frac{3}{10}} w}{371450 \pi } \Big(-\frac{910}{11} \sqrt{3}
  c_1^3+\frac{2240 c_1^3}{33 \sqrt{3}}+\frac{425}{11} \sqrt{35} c_0
  c_2 c_1-77 \sqrt{30} c_{-1} c_3 c_1-7 \sqrt{\frac{15}{2}} c_{-1} c_3
  c_1+\frac{907}{11} \sqrt{10} c_{-2} c_4 c_1-\\ 24 \sqrt{55} c_{-3}
  c_5 c_1+300 \sqrt{\frac{2}{11}} c_{-4} c_6 c_1-\frac{1814}{11}
  \sqrt{3} c_{-1} c_2^2+\frac{4172 c_{-1} c_2^2}{11
    \sqrt{3}}+\frac{257}{11} \sqrt{30} c_{-3} c_3^2-111
  \sqrt{\frac{15}{2}} c_{-3} c_3^2-42 \sqrt{22} c_{-5} c_4^2-\\ 71
  \sqrt{\frac{11}{2}} c_{-5} c_4^2+450 \sqrt{\frac{2}{11}}
  c_{-5}c_4^2+\frac{280}{11} \sqrt{30} c_0^2 c_3-7 \sqrt{\frac{15}{2}}
  c_0^2 c_3+\frac{686}{11} \sqrt{30} c_{-2} c_2 c_3+\frac{5}{11}
  \sqrt{42} c_{-1} c_0 c_4-5 \sqrt{\frac{21}{2}} c_{-1} c_0
  c_4-\\ \frac{3990}{11} c_{-3} c_2 c_4+\frac{1363}{11} \sqrt{30}
  c_{-4} c_3 c_4-111 \sqrt{\frac{15}{2}} c_{-4} c_3 c_4-\frac{105
    c_{-1}^2 c_5}{\sqrt{22}}-175 \sqrt{\frac{2}{11}} c_{-1}^2 c_5+30
  \sqrt{\frac{210}{11}} c_{-2} c_0 c_5+\\ 221 \sqrt{\frac{15}{11}}
  c_{-4} c_2 c_5-20 \sqrt{30} c_{-5} c_3 c_5-71 \sqrt{\frac{15}{2}}
  c_{-5} c_3 c_5+155 \sqrt{3} c_{-6} c_4 c_5+56 \sqrt{\frac{15}{11}}
  c_{-2} c_{-1} c_6-48 \sqrt{\frac{70}{11}} c_{-3} c_0 c_6-\\ 33
  \sqrt{10} c_{-5} c_2 c_6+11 \sqrt{30} c_{-6} c_3
  c_6\Big)+\frac{122499 \sqrt{\frac{3}{10}} w}{742900 \pi }
  \Big(-\frac{1260}{11} \sqrt{3} c_1^3+\frac{1120 c_1^3}{11
    \sqrt{3}}+\frac{570}{11} \sqrt{35} c_0 c_2 c_1-\frac{1197}{11}
  \sqrt{30} c_{-1} c_3 c_1+\\ \frac{1218}{11} \sqrt{10} c_{-2} c_4
  c_1-336 \sqrt{\frac{5}{11}} c_{-3} c_5 c_1+420 \sqrt{\frac{2}{11}}
  c_{-4} c_6 c_1-\frac{560}{11} \sqrt{3} c_{-1} c_2^2-\frac{469}{11}
  \sqrt{30} c_{-3} c_3^2-105 \sqrt{22} c_{-5}
  c_4^2+\\ 630\sqrt{\frac{2}{11}} c_{-5} c_4^2+\frac{326}{11}
  \sqrt{30} c_0^2 c_3+84 \sqrt{30} c_{-2} c_2 c_3-\frac{25}{11}
  \sqrt{42} c_{-1} c_0 c_4-\frac{5460}{11} c_{-3} c_2
  c_4+\frac{1015}{11} \sqrt{30} c_{-4} c_3 c_4-315 \sqrt{\frac{2}{11}}
  c_{-1}^2 c_5+\\ 40 \sqrt{\frac{210}{11}} c_{-2} c_0 c_5+294
  \sqrt{\frac{15}{11}} c_{-4} c_2 c_5-77 \sqrt{30} c_{-5} c_3 c_5+210
  \sqrt{3} c_{-6} c_4 c_5+84 \sqrt{\frac{15}{11}} c_{-2} c_{-1} c_6-72
  \sqrt{\frac{70}{11}} c_{-3} c_0 c_6-\\ 42 \sqrt{10} c_{-5} c_2
  c_6+14 \sqrt{30} c_{-6} c_3 c_6\Big) =0
\end{multline}
\clearpage
\begin{multline}
  G_2:=r c_2-\frac{10 \sqrt{\frac{13}{231 \pi }} u}{3553}
  \Big(-42\sqrt{55} c_1^2+22 \sqrt{231} c_0 c_2+ 7 \Big(9 \sqrt{22}
  c_{-1} c_3-12 \sqrt{66} c_{-2} c_4+22 \sqrt{3} c_{-3} c_5+22
  \sqrt{30} c_{-4} c_6\Big)\Big)\\ +\frac{122499 w}{1485800 \pi }
  \Big(168 \sqrt{\frac{5}{11}} c_6 c_{-2}^2+\frac{4424}{11} c_2^2
  c_{-2}+\frac{672}{11} \sqrt{10} c_1 c_3 c_{-2}+\frac{1608}{11}
  \sqrt{14} c_0 c_4 c_{-2}-56 \sqrt{\frac{6}{11}} c_{-1} c_5
  c_{-2}-\\ \frac{152}{11} \sqrt{105} c_0 c_1^2+\frac{546}{11}
  \sqrt{30} c_{-4} c_3^2+252 \sqrt{\frac{5}{11}} c_{-6}
  c_4^2+\frac{5946}{11} c_0^2 c_2-\frac{5432}{11} c_{-1} c_1
  c_2-\frac{570}{11} \sqrt{42} c_{-1} c_0 c_3-504 c_{-3} c_2
  c_3-\\ \frac{266}{11} \sqrt{30} c_{-1}^2 c_4-\frac{2436}{11}
  \sqrt{3} c_{-3} c_1 c_4+\frac{3864}{11} c_{-4} c_2 c_4-287 \sqrt{22}
  c_{-5} c_3 c_4+2275 \sqrt{\frac{2}{11}} c_{-5} c_3 c_4-240
  \sqrt{\frac{7}{11}} c_{-3} c_0 c_5+\\ 420 \sqrt{\frac{5}{11}} c_{-4}
  c_1 c_5-252 c_{-5} c_2 c_5+84 \sqrt{3} c_{-6} c_3 c_5-252
  \sqrt{\frac{2}{11}} c_{-3} c_{-1} c_6+12 \sqrt{\frac{70}{11}} c_{-4}
  c_0 c_6\Big)-\\ \frac{124033 w}{742900 \pi } \Big(112
  \sqrt{\frac{5}{11}} c_6 c_{-2}^2+\frac{9808}{33} c_2^2
  c_{-2}+\frac{508}{11} \sqrt{10} c_1 c_3 c_{-2}+\frac{1192}{11}
  \sqrt{14} c_0 c_4 c_{-2}+266 \sqrt{\frac{6}{11}} c_{-1} c_5
  c_{-2}-910 \sqrt{\frac{2}{33}} c_{-1} c_5 c_{-2}-\\ \frac{546}{11}
  \sqrt{105} c_0 c_1^2+118 \sqrt{\frac{35}{3}} c_0
  c_1^2+\frac{399}{11} \sqrt{30} c_{-4} c_3^2+180 \sqrt{\frac{5}{11}}
  c_{-6} c_4^2+\frac{4424}{11} c_0^2 c_2-\frac{12094}{33} c_{-1} c_1
  c_2-\frac{425}{11} \sqrt{42} c_{-1} c_0 c_3-\\ \frac{4116}{11}
  c_{-3} c_2 c_3-\frac{70}{11} \sqrt{30} c_{-1}^2 c_4-35
  \sqrt{\frac{10}{3}} c_{-1}^2 c_4-\frac{1814}{11} \sqrt{3} c_{-3} c_1
  c_4+\frac{2956}{11} c_{-4} c_2 c_4-266 \sqrt{22} c_{-5} c_3 c_4+2263
  \sqrt{\frac{2}{11}} c_{-5} c_3 c_4-\\ 24 \sqrt{77} c_{-3} c_0 c_5+84
  \sqrt{\frac{7}{11}} c_{-3} c_0 c_5-24 \sqrt{55} c_{-4} c_1 c_5+562
  \sqrt{\frac{5}{11}} c_{-4} c_1 c_5-178 c_{-5} c_2 c_5+66 \sqrt{3}
  c_{-6} c_3 c_5-168 \sqrt{\frac{2}{11}} c_{-3} c_{-1} c_6+\\ 12
  \sqrt{\frac{70}{11}} c_{-4} c_0 c_6-2 \sqrt{30} c_{-5} c_1
  c_6\Big)+\frac{6461 w}{37145 \pi } \Big(28 \sqrt{\frac{5}{11}} c_6
  c_{-2}^2+\frac{3370}{33} c_2^2 c_{-2}+\frac{172}{11} \sqrt{10} c_1
  c_3 c_{-2}+\frac{388}{11} \sqrt{14} c_0 c_4 c_{-2}+\\ 28
  \sqrt{\frac{22}{3}} c_{-1} c_5 c_{-2}-112 \sqrt{\frac{6}{11}} c_{-1}
  c_5 c_{-2}-\frac{168}{11} \sqrt{105} c_0 c_1^2+\frac{392}{11}
  \sqrt{\frac{35}{3}} c_0 c_1^2+\frac{126}{11} \sqrt{30} c_{-4}
  c_3^2+54 \sqrt{\frac{5}{11}} c_{-6} c_4^2+\frac{1484}{11} c_0^2
  c_2-\\ \frac{4144}{33} c_{-1} c_1 c_2-\frac{140}{11} \sqrt{42}
  c_{-1} c_0 c_3-\frac{1410}{11} c_{-3} c_2 c_3-\frac{14}{11}
  \sqrt{30} c_{-1}^2 c_4-14 \sqrt{\frac{10}{3}} c_{-1}^2
  c_4-\frac{596}{11} \sqrt{3} c_{-3} c_1 c_4+\frac{1090}{11} c_{-4}
  c_2 c_4-\\ 68 \sqrt{22} c_{-5} c_3 c_4+526 \sqrt{\frac{2}{11}}
  c_{-5} c_3 c_4+16 \sqrt{77} c_{-3} c_0 c_5-236 \sqrt{\frac{7}{11}}
  c_{-3} c_0 c_5+8 \sqrt{55} c_{-4} c_1 c_5-58 c_{-5} c_2 c_5+24
  \sqrt{3} c_{-6} c_3 c_5-\\ 42 \sqrt{\frac{2}{11}} c_{-3} c_{-1}
  c_6+6 \sqrt{\frac{70}{11}} c_{-4} c_0 c_6-2 \sqrt{30} c_{-5} c_1
  c_6+6 c_{-6} c_2 c_6\Big) =0
\end{multline}
\clearpage
\begin{multline}
  G_1:=r c_1+\frac{\sqrt{\frac{130}{231 \pi }} u}{3553} \Big(20
  \sqrt{2310} c_0 c_1-7 \Big(60 \sqrt{22} c_{-1} c_2-18 \sqrt{55}
  c_{-2} c_3-25 \sqrt{66} c_{-3} c_4+99 \sqrt{10} c_{-4} c_5+22
  \sqrt{165} c_{-5} c_6\Big)\Big)\\ +\frac{19383 w}{37145 \sqrt{10}
    \pi } \Big(\frac{56}{3} \sqrt{\frac{55}{3}} c_5 c_{-2}^2-196
  \sqrt{\frac{5}{33}} c_5 c_{-2}^2+\frac{4144}{99} \sqrt{10} c_1 c_2
  c_{-2}-\frac{36}{11} \sqrt{105} c_0 c_3 c_{-2}+\frac{388}{11}
  \sqrt{\frac{35}{3}} c_0 c_3 c_{-2}+\\ \frac{3920 c_{-1} c_4
    c_{-2}}{33 \sqrt{3}}+28 \sqrt{\frac{5}{11}} c_{-3} c_6
  c_{-2}-\frac{5320}{99} \sqrt{10} c_{-1} c_1^2-\frac{860}{33} c_{-3}
  c_2^2-68 \sqrt{\frac{55}{3}} c_{-5} c_3^2+42 \sqrt{\frac{15}{11}}
  c_{-5} c_3^2+526 \sqrt{\frac{5}{33}} c_{-5} c_3^2+\\ \frac{700}{33}
  \sqrt{10} c_0^2 c_1-\frac{140}{11} \sqrt{42} c_{-1} c_0
  c_2+\frac{2380}{33} \sqrt{\frac{14}{3}} c_{-1} c_0
  c_2-\frac{3920}{33} c_{-1}^2 c_3-\frac{596}{11} \sqrt{10} c_{-3} c_1
  c_3+\frac{17}{11} \sqrt{30} c_{-4} c_2 c_3+\frac{545}{11}
  \sqrt{\frac{10}{3}} c_{-4} c_2 c_3-\\ \frac{20}{11} \sqrt{14} c_{-3}
  c_0 c_4+\frac{680}{33} \sqrt{10} c_{-4} c_1 c_4-\frac{40}{3}
  \sqrt{22} c_{-5} c_2 c_4+4 \sqrt{66} c_{-6} c_3 c_4-68
  \sqrt{\frac{22}{3}} c_{-6} c_3 c_4+90 \sqrt{\frac{6}{11}} c_{-6} c_3
  c_4+526 \sqrt{\frac{2}{33}} c_{-6} c_3 c_4+\\ 28 \sqrt{66} c_{-3}
  c_{-1} c_5-126 \sqrt{\frac{6}{11}} c_{-3} c_{-1} c_5-826
  \sqrt{\frac{2}{33}} c_{-3} c_{-1} c_5+4 \sqrt{2310} c_{-4} c_0 c_5+2
  \sqrt{\frac{210}{11}} c_{-4} c_0 c_5-118 \sqrt{\frac{70}{33}} c_{-4}
  c_0 c_5-\\ \frac{20}{3} \sqrt{10} c_{-5} c_1 c_5+26 \sqrt{3} c_{-6}
  c_2 c_5-\frac{58 c_{-6} c_2 c_5}{\sqrt{3}}\Big)+\frac{367497
    w}{1485800 \sqrt{10} \pi } \Big(56 \sqrt{\frac{15}{11}} c_5
  c_{-2}^2-112 \sqrt{\frac{5}{33}} c_5 c_{-2}^2+\frac{5432}{33}
  \sqrt{10} c_1 c_2 c_{-2}+\\ \frac{380}{11} \sqrt{105} c_0 c_3
  c_{-2}+\frac{3752}{11} \sqrt{3} c_{-1} c_4 c_{-2}-\frac{5936 c_{-1}
    c_4 c_{-2}}{11 \sqrt{3}}+168 \sqrt{\frac{5}{11}} c_{-3} c_6
  c_{-2}-\frac{7028}{33} \sqrt{10} c_{-1} c_1^2-\frac{1120}{11} c_{-3}
  c_2^2-\\ 287 \sqrt{\frac{55}{3}} c_{-5} c_3^2+182
  \sqrt{\frac{15}{11}} c_{-5} c_3^2+2275 \sqrt{\frac{5}{33}} c_{-5}
  c_3^2+\frac{918}{11} \sqrt{10} c_0^2 c_1+\frac{1412}{11} \sqrt{42}
  c_{-1} c_0 c_2-\frac{2716}{11} \sqrt{\frac{14}{3}} c_{-1} c_0
  c_2-\frac{5320}{11} c_{-1}^2 c_3-\\ \frac{2394}{11} \sqrt{10} c_{-3}
  c_1 c_3+\frac{812}{11} \sqrt{30} c_{-4} c_2 c_3-\frac{50}{11}
  \sqrt{14} c_{-3} c_0 c_4+\frac{840}{11} \sqrt{10} c_{-4} c_1 c_4-329
  \sqrt{22} c_{-5} c_2 c_4+2919 \sqrt{\frac{2}{11}} c_{-5} c_2
  c_4+\\ 14 \sqrt{66} c_{-6} c_3 c_4-287 \sqrt{\frac{22}{3}} c_{-6}
  c_3 c_4+420 \sqrt{\frac{6}{11}} c_{-6} c_3 c_4+2275
  \sqrt{\frac{2}{33}} c_{-6} c_3 c_4-420 \sqrt{\frac{6}{11}} c_{-3}
  c_{-1} c_5+34 \sqrt{\frac{210}{11}} c_{-4} c_0 c_5-\\ 14 \sqrt{10}
  c_{-5} c_1 c_5-56 \sqrt{\frac{15}{11}} c_{-4} c_{-1} c_6+4 \sqrt{35}
  c_{-5} c_0 c_6\Big)-\frac{372099 w}{742900 \sqrt{10} \pi}
  \Big(\frac{112}{3} \sqrt{\frac{5}{33}} c_5 c_{-2}^2+\frac{12094}{99}
  \sqrt{10} c_1 c_2 c_{-2}-\\ \frac{114}{11} \sqrt{105} c_0 c_3
  c_{-2}+\frac{1192}{11} \sqrt{\frac{35}{3}} c_0 c_3
  c_{-2}+\frac{11900 c_{-1} c_4 c_{-2}}{33 \sqrt{3}}+112
  \sqrt{\frac{5}{11}} c_{-3} c_6 c_{-2}-\frac{1424}{9} \sqrt{10}
  c_{-1} c_1^2-\frac{2540}{33} c_{-3} c_2^2-\\ 266 \sqrt{\frac{55}{3}}
  c_{-5} c_3^2+133 \sqrt{\frac{15}{11}} c_{-5} c_3^2+2263
  \sqrt{\frac{5}{33}} c_{-5} c_3^2+\frac{2044}{33} \sqrt{10} c_0^2
  c_1-\frac{425}{11} \sqrt{42} c_{-1} c_0 c_2+\frac{7225}{33}
  \sqrt{\frac{14}{3}} c_{-1} c_0 c_2-\\ \frac{11900}{33} c_{-1}^2
  c_3-161 \sqrt{10} c_{-3} c_1 c_3+\frac{112}{11} \sqrt{30} c_{-4} c_2
  c_3+\frac{1478}{11} \sqrt{\frac{10}{3}} c_{-4} c_2 c_3-\frac{45}{11}
  \sqrt{14} c_{-3} c_0 c_4+\frac{1874}{33} \sqrt{10} c_{-4} c_1
  c_4-\\ \frac{887}{3} \sqrt{22} c_{-5} c_2 c_4+\frac{8267}{3}
  \sqrt{\frac{2}{11}} c_{-5} c_2 c_4+11 \sqrt{66} c_{-6} c_3 c_4-266
  \sqrt{\frac{22}{3}} c_{-6} c_3 c_4+300 \sqrt{\frac{6}{11}} c_{-6}
  c_3 c_4+2263 \sqrt{\frac{2}{33}} c_{-6} c_3 c_4-\\ 910
  \sqrt{\frac{2}{33}} c_{-3} c_{-1} c_5-8 \sqrt{2310} c_{-4} c_0
  c_5+18 \sqrt{\frac{210}{11}} c_{-4} c_0 c_5+281 \sqrt{\frac{70}{33}}
  c_{-4} c_0 c_5-\frac{35}{3} \sqrt{10} c_{-5} c_1 c_5+66 \sqrt{3}
  c_{-6} c_2 c_5-\\ \frac{178 c_{-6} c_2 c_5}{\sqrt{3}}-28
  \sqrt{\frac{15}{11}} c_{-4} c_{-1} c_6+2 \sqrt{35} c_{-5} c_0 c_6-2
  \sqrt{10} c_{-6} c_1 c_6\Big) =0
\end{multline}
\clearpage
\begin{multline}
  G_0:=r c_0+\frac{10 \sqrt{\frac{13}{\pi }} u}{3553} \Big(20 c_0^2-20
  c_{-1} c_1-22 c_{-2} c_2+43 c_{-3} c_3-8 c_{-4} c_4-55 c_{-5} c_5-22
  c_{-6} c_6\Big)+\\ \frac{923 \sqrt{42} w}{37145 \pi }
  \Big(\frac{350}{11} \sqrt{\frac{14}{3}} c_0^3-\frac{140}{11}
  \sqrt{42} c_{-1} c_1 c_0-\frac{280}{11} \sqrt{\frac{14}{3}} c_{-1}
  c_1 c_0+\frac{1484}{11} \sqrt{\frac{14}{3}} c_{-2} c_2
  c_0-\frac{168}{11} \sqrt{42} c_{-3} c_3 c_0+\\ \frac{140}{11}
  \sqrt{42} c_{-4} c_4 c_0-\frac{392}{33} \sqrt{10} c_{-2}
  c_1^2+\frac{596}{11} \sqrt{3} c_{-4} c_2^2-\frac{430 c_{-4}
    c_2^2}{11 \sqrt{3}}-62 \sqrt{22} c_{-6} c_3^2+724
  \sqrt{\frac{2}{11}} c_{-6} c_3^2-\frac{392}{33} \sqrt{10} c_{-1}^2
  c_2-\frac{980}{11} c_{-3} c_1 c_2-\\ \frac{980}{11} c_{-2} c_{-1}
  c_3+\frac{14}{11} \sqrt{30} c_{-4} c_1 c_3-68 \sqrt{66} c_{-5} c_2
  c_3+\frac{545 c_{-5} c_2 c_3}{\sqrt{66}}-20 \sqrt{\frac{22}{3}}
  c_{-5} c_2 c_3+652 \sqrt{\frac{6}{11}} c_{-5} c_2 c_3+17
  \sqrt{\frac{3}{22}} c_{-5} c_2 c_3+\\ \frac{1358 c_{-2}^2 c_4}{11
    \sqrt{3}}-\frac{14}{11} \sqrt{30} c_{-3} c_{-1} c_4+\frac{392}{11}
  \sqrt{\frac{10}{3}} c_{-3} c_{-1} c_4+91 \sqrt{\frac{6}{5}} c_{-3}
  c_{-1} c_4-413 \sqrt{\frac{2}{15}} c_{-3} c_{-1} c_4-\frac{50}{3}
  \sqrt{22} c_{-5} c_1 c_4+\\ \frac{340}{3} \sqrt{\frac{2}{11}} c_{-5}
  c_1 c_4-8 \sqrt{\frac{55}{3}} c_{-6} c_2 c_4-43 \sqrt{\frac{33}{5}}
  c_{-6} c_2 c_4-29 \sqrt{\frac{11}{15}} c_{-6} c_2 c_4+796
  \sqrt{\frac{3}{55}} c_{-6} c_2 c_4+56 \sqrt{\frac{22}{3}} c_{-3}
  c_{-2} c_5-\\ 826 \sqrt{\frac{2}{33}} c_{-3} c_{-2} c_5+84 \sqrt{22}
  c_{-4} c_{-1} c_5-994 \sqrt{\frac{2}{11}} c_{-4} c_{-1} c_5+42
  \sqrt{\frac{2}{11}} c_{-3}^2 c_6+14 \sqrt{\frac{15}{11}} c_{-4}
  c_{-2} c_6\Big)-\\ \frac{17719 \sqrt{\frac{21}{2}} w}{371450 \pi }
  \Big(\frac{1022}{11} \sqrt{\frac{14}{3}} c_0^3-\frac{425}{11}
  \sqrt{42} c_{-1} c_1 c_0-\frac{769}{11} \sqrt{\frac{14}{3}} c_{-1}
  c_1 c_0+\frac{4424}{11} \sqrt{\frac{14}{3}} c_{-2} c_2
  c_0-\frac{483}{11} \sqrt{42} c_{-3} c_3 c_0-\\ 35 \sqrt{42} c_{-4}
  c_4 c_0-\frac{45}{11} \sqrt{\frac{21}{2}} c_{-4} c_4
  c_0+\frac{937}{11} \sqrt{\frac{14}{3}} c_{-4} c_4 c_0+281
  \sqrt{\frac{7}{6}} c_{-4} c_4 c_0+7 \sqrt{42} c_{-5} c_5
  c_0-\frac{1190}{33} \sqrt{10} c_{-2} c_1^2+\frac{1814}{11} \sqrt{3}
  c_{-4} c_2^2-\\ \frac{1270 c_{-4} c_2^2}{11 \sqrt{3}}-12 \sqrt{22}
  c_{-6} c_3^2+\frac{3163 c_{-6} c_3^2}{\sqrt{22}}-233
  \sqrt{\frac{11}{2}} c_{-6} c_3^2-\frac{1190}{33} \sqrt{10} c_{-1}^2
  c_2-\frac{2975}{11} c_{-3} c_1 c_2-\frac{2975}{11} c_{-2} c_{-1}
  c_3+\\ \frac{917}{11} \sqrt{30} c_{-4} c_1 c_3-161
  \sqrt{\frac{15}{2}} c_{-4} c_1 c_3-24 \sqrt{66} c_{-5} c_2
  c_3+\frac{8267 c_{-5} c_2 c_3}{\sqrt{66}}-887 \sqrt{\frac{11}{6}}
  c_{-5} c_2 c_3+56 \sqrt{\frac{6}{11}} c_{-5} c_2 c_3+\\ 739
  \sqrt{\frac{2}{33}} c_{-5} c_2 c_3+\frac{4172 c_{-2}^2 c_4}{11
    \sqrt{3}}+\frac{63}{11} \sqrt{30} c_{-3} c_{-1} c_4-\frac{63}{11}
  \sqrt{\frac{15}{2}} c_{-3} c_{-1} c_4-\frac{887}{3} \sqrt{22} c_{-5}
  c_1 c_4-\frac{35}{3} \sqrt{\frac{11}{2}} c_{-5} c_1 c_4+\\ 3068
  \sqrt{\frac{2}{11}} c_{-5} c_1 c_4-40 \sqrt{165} c_{-6} c_2
  c_4+\frac{8267 c_{-6} c_2 c_4}{\sqrt{165}}-976 \sqrt{\frac{11}{15}}
  c_{-6} c_2 c_4+3163 \sqrt{\frac{3}{55}} c_{-6} c_2 c_4-210
  \sqrt{\frac{6}{11}} c_{-3} c_{-2} c_5-\\ 84 \sqrt{22} c_{-4} c_{-1}
  c_5+\frac{2345 c_{-4} c_{-1} c_5}{\sqrt{22}}-497 \sqrt{\frac{2}{11}}
  c_{-4} c_{-1} c_5-7 \sqrt{3} c_{-6} c_1 c_5+168 \sqrt{\frac{2}{11}}
  c_{-3}^2 c_6+28 \sqrt{\frac{15}{11}} c_{-4} c_{-2} c_6-7 \sqrt{3}
  c_{-5} c_{-1} c_6\Big)+\\ \frac{122499 \sqrt{\frac{3}{14}}
    w}{742900\pi } \Big(\frac{459}{11} \sqrt{42} c_0^3-\frac{918}{11}
  \sqrt{42} c_{-1} c_1 c_0+\frac{1982}{11} \sqrt{42} c_{-2} c_2
  c_0-\frac{652}{11} \sqrt{42} c_{-3} c_3 c_0+\frac{582}{11} \sqrt{42}
  c_{-4} c_4 c_0+12 \sqrt{42} c_{-5} c_5 c_0+\\ 2 \sqrt{42} c_{-6} c_6
  c_0-\frac{532}{11} \sqrt{10} c_{-2} c_1^2+\frac{1876}{11} \sqrt{3}
  c_{-4} c_2^2-266 \sqrt{22} c_{-6} c_3^2+3178 \sqrt{\frac{2}{11}}
  c_{-6} c_3^2-\frac{532}{11} \sqrt{10} c_{-1}^2 c_2-\frac{3990}{11}
  c_{-3} c_1 c_2-\\ \frac{3990}{11} c_{-2} c_{-1} c_3+\frac{35}{11}
  \sqrt{30} c_{-4} c_1 c_3-287 \sqrt{66} c_{-5} c_2 c_3-329
  \sqrt{\frac{33}{2}} c_{-5} c_2 c_3+3227 \sqrt{\frac{6}{11}} c_{-5}
  c_2 c_3+2919 \sqrt{\frac{3}{22}} c_{-5} c_2 c_3+\\ \frac{1876}{11}
  \sqrt{3} c_{-2}^2 c_4+\frac{35}{11} \sqrt{30} c_{-3} c_{-1} c_4-336
  \sqrt{22} c_{-5} c_1 c_4+3339 \sqrt{\frac{2}{11}} c_{-5} c_1 c_4-49
  \sqrt{165} c_{-6} c_2 c_4-329 \sqrt{\frac{33}{5}} c_{-6} c_2
  c_4+\\ 707 \sqrt{\frac{15}{11}} c_{-6} c_2 c_4+2919
  \sqrt{\frac{3}{55}} c_{-6} c_2 c_4-280 \sqrt{\frac{6}{11}} c_{-3}
  c_{-2} c_5-357 \sqrt{\frac{2}{11}} c_{-4} c_{-1} c_5-14 \sqrt{3}
  c_{-6} c_1 c_5+252 \sqrt{\frac{2}{11}} c_{-3}^2 c_6+\\ 28
  \sqrt{\frac{15}{11}} c_{-4} c_{-2} c_6-14 \sqrt{3} c_{-5} c_{-1}
  c_6\Big) =0
\end{multline}
\section{References}
\bibliography{mybib}